\documentclass[a4paper,fleqn,usenatbib]{mnras}

\usepackage{newtxtext,newtxmath}
\usepackage[T1]{fontenc}
\usepackage{ae,aecompl}
\usepackage{subfigure}

\usepackage{graphicx}
\usepackage{float}
\usepackage{placeins} 
\usepackage{paralist}

\usepackage{todonotes}

\newcommand{\HI}{H\,{\sc i}}

\newcommand{\kms}{km\,s$^{-1}$}

\title[Galactic HI foreground of GRBs]{Exploring the pattern of the Galactic HI foreground of GRBs with the ATCA}

\author[H. D\'{e}nes et al.]{
	H. D\'{e}nes$^{1}$,\thanks{E-mail: denes@astron.nl (HD)}
	P.A. Jones$^{2}$,
	L.V. T\'{o}th$^{3}$,
	S. Zahorecz$^{4,5}$,
	B-C. Koo$^{6}$,
	S. Pinter$^{3,7}$,
	\newauthor
	I.I. Racz$^{3,7,8}$,
	L.G. Bal\'{a}zs$^{3,8}$, 
	M.R. Cunningham$^{2}$,
	Y. Doi$^{9}$,
	I. Horvath$^{7}$,
	T. Kov\'{a}cs$^{3}$,
	\newauthor
	T. Onishi$^{4}$,
	N. Suleiman$^{3}$,
	Z. Bagoly$^{3,8}$
	\\
	$^{1}$ASTRON, The Netherlands Institute for Radio Astronomy, Dwingeloo, the Netherlands \\
	$^{2}$School of Physics, University of New South Wales, Sydney, NSW, 2052, Australia\\
	$^{3}$Department of Astronomy of the  E\"otv\"os Lor\'and University,
	P\'azm\'any P\'eter s\'et\'any 1, 1117 Budapest, Hungary\\
	$^{4}$Department of Physical Science, Graduate School of Science, Osaka Prefecture University, 1-1 Gakuen-cho, Naka-ku, Sakai, \\ Osaka 599-8531, Japan\\
	$^{5}$National Astronomical Observatory of Japan, National Institutes of Natural Science, 2-21-1 Osawa, Mitaka, Tokyo 181-8588, Japan\\
	$^{6}$Seoul National University, Korea\\
	$^{7}$National University of Public Service, Budapest, Hungary\\ 
	$^{8}$Konkoly Observatory of the Hungarian Academy of Sciences, H-1121 Budapest, Konkoly Thege Mikl\'os \'ut 15-17.\\
	$^{9}$Department of Earth Science and Astronomy, The University of Tokyo, 3-8-1 Komaba, Meguro-ku,Tokyo 153-8902, Japan\\	
}

\date{Accepted XXX. Received YYY; in original form ZZZ}

\pubyear{2018}

\begin{document}
\label{firstpage}
\pagerange{\pageref{firstpage}--\pageref{lastpage}}
\maketitle
	
\begin{abstract}
The afterglow of a gamma ray burst (GRB) can give us valuable insight into the properties of its host galaxy. To correctly interpret the spectra of the afterglow we need to have a good understanding of the foreground interstellar medium (ISM) in our own Galaxy. The common practice to correct for the foreground is to use neutral hydrogen (\HI) data from the Leiden/Argentina/Bonn (LAB) survey. However, the poor spatial resolution of the single dish data may have a significant effect on the derived column densities. To investigate this, we present new high-resolution \HI\ observations with the Australia Telescope Compact Array (ATCA) towards 4 GRBs. We combine the interferometric ATCA data with single dish data from the Galactic All Sky Survey (GASS) and derive new Galactic \HI\ column densities towards the GRBs. We use these new foreground column densities to fit the Swift XRT X-ray spectra and calculate new intrinsic hydrogen column density values for the GRB host galaxies. We find that the new ATCA data shows higher Galactic \HI\ column densities compared to the previous single dish data, which results in lower intrinsic column densities for the hosts. We investigate the line of sight optical depth near the GRBs and find that it may not be negligible towards one of the GRBs, which indicates that the intrinsic hydrogen column density of its host galaxy may be even lower. In addition, we compare our results to column densities derived from far-infrared data and find a reasonable agreement with the \HI\ data.  
\end{abstract}
	
\begin{keywords}
	ISM: structure -- galaxies: general, ISM -- radio lines: ISM -- gamma rays: bursts.
\end{keywords}
	
	
	
\section{Introduction}
	
Gamma-ray bursts (GRBs) are sudden, intense flashes of gamma rays that last for a few seconds and are typically followed by an afterglow. In the last decades several major advances have been made regarding GRBs, including the discovery of slowly fading X-ray (e.g. \citealt{theory_afterglow,theory_afterglow_2,theory_afterglow_3,afterglow_xray}), optical (e.g. \citealt{theory_afterglow_4,vanParadijs1997}), and radio (e.g. \citealt{afterglow_radio}) afterglows; the prompt high energy (100 MeV to GeV) emission of the GRBs (e.g. \citealt{Schneid1992, Hurley1994, Abdo_2009, Ghirlanda2010}); identification of host galaxies at cosmological distances (e.g. \citealt{Metzger1997,Kulkarni1998}); and evidence showing that many long-duration GRBs are associated with supernovae and star-forming regions (e.g. \citealt{Galama1998, Stanek2003, Hjorth2003}). Considering that GRBs can be detected as far as redshift $\sim$8 (\citealt{Tanvir2009, Salvaterra2009, Cucchiara2011}), they can be valuable probes of star formation in the distant Universe.  
		
One of the main sources of information on GRBs is their optical \citep{eliasdottir2009} and X-ray afterglow spectra, which can give us direct information on the circumstellar and interstellar medium in the immediate environment of the GRB progenitor (e.g. \citealt{Schady2015} and references therein). The extinction seen in the X-ray and optical spectra of the afterglows (eg. \citealt{Fynbo2009, Watson2013, Cucchiara2015}) can be modelled and the intrinsic hydrogen column density of the host galaxy ($N($H$)_\mathrm{GRB}$) can be derived. However, the afterglow radiation also passes trough the intergalactic medium (IGM) and the Galactic Interstellar Medium (ISM) before it is detected, which adds another level of complexity. For example, there is a discrepancy between the column densities derived from UV/optical spectroscopy and X-ray absorption. Column densities from X-ray absorption are generally higher, even after accounting for the foreground absorption by the Milky Way (e.g. \citealt{Galama2001, Watson2007, Campana2010, Schady2011, Heintz2018}). This can have several reasons, starting from not probing the same gas phases and spatial regions of the host galaxy to several other effects. UV and optical spectroscopy probes distinct gas phases in the host galaxies general ISM while X-ray absorption probes the total column density of gas along the line of sight, including the cold, warm and hot phases (e.g. \citealt{Schady2015}). Additional effects can be absorption by the intergalactic medium (e.g. \citealt{Campana2010, Campana2012, Starling2013}), absorption by ultra ionised gas near the GRB \citep{Schady2011}, the GRB being embedded in a dense and turbulent ISM (e.g. \citealt{Krongold2013, Tanga2016}) or even underestimating the Milky Way foreground (e.g. \citealt{Wilms2000}).

To better understand the properties of GRBs and their host galaxies, several systematic GRB host galaxy surveys were carried out in recent years. Such surveys were: the Optically Unbiased Gamma-Ray Burst Host (TOUGH; \citealt{Hjorth2012}) survey, with a sample of 69 GRB host galaxies selected based on their X-ray properties; the BAT6 survey (\citealt{Salvaterra2012}) with a sample of the 58 brightest GRBs observed by \textit{Swift} at the time; and the Swift Gamma-Ray Burst Host Galaxy Legacy Survey (SHOALS; \citealt{Perley2016}), a homogeneously targeted survey of 119 GRB host galaxies. These surveys provided large samples of GRB host galaxies with redshifts and derived intrinsic properties such as stellar masses and star formation rates.

In this paper, we are focusing on estimating the Milky Way foreground($N($H$)_\mathrm{MW}$). There are various ways to estimate $N($H$)_\mathrm{MW}$: 
\begin{inparaenum} 
	\item combining galaxy counts and \HI\ measurements \citep{burstein1982}; 
	\item deriving extinction from the Galactic \HI\ surveys such as the Leiden/Argentine/Bonn (LAB) survey \citep{kalberla2005}; 
	\item using extinction maps calculated from infrared surveys (e.g. \citealt{Schlegel1998,schlafly2011}); or
	\item from spectroscopic measurements and colours of nearby Galactic stars (e.g. \citealt{Green2018} and references therein).
\end{inparaenum} 
The standard procedure applied by the UK Swift Science Data Centre\footnote{\url{http://www.swift.ac.uk/}} (UKSSDC) to take into account the contribution of the Galactic foreground is to use the Galactic \HI\ data from the LAB survey. However, the spatial resolution of the LAB data is 36', which is about a hundred times the size of the Swift position accuracy ($\sim$ 2.4"). In addition to this, the spatial distribution of the ISM is far from a smooth distribution. Filamentary structures were discovered from arc minute to arc second scales (e.g. \citealt{McClure-Griffiths2006, Clark2014, Kalberla2016}) in the Parkes Galactic All Sky Survey (GASS, \citealt{McClure-Griffiths2009, Kalberla2010, Kalberla2015}), HI4PI \citep{hi4pi2016}, the Galactic Arecibo L-Band Feed Array \HI\ Survey (GALFA-\HI; \citealt{Peek2011}) and the SGPS Galactic Centre Survey (SGPS GC; \citealt{McClure-Griffiths2012}). Consequently, the estimated surface density of the  foreground matter strongly depends on the beam width of the survey. Fine ISM structures in the line of sight of the GRBs could result in an over or under estimation of the true value when using a large beam width. To study this effect, observations with smaller beam widths are highly desirable to get more reliable estimates of the foreground matter and, consequently, more accurate intrinsic data on GRBs and their close environment.
	
We note, that there are already \HI\ data available from all sky surveys with better spatial resolution. For example, the HI4PI data \citep{hi4pi2016} with a spatial resolution of 16.5'. Beside the \HI\ data, far-infrared (FIR) all sky surveys can also be used to estimate the optical and X-ray extinction assuming an appropriate dust model. After the IRAS based maps (\citealt{Schlegel1998,schlafly2011}), recently higher resolution reddening and extinction maps were derived from the Planck FIR and sub-mm all sky survey data (\citealt{planck2014,planck_dust}). These new maps have a typical angular resolution of 6'.
	
\cite{Toth2017} investigated the effect of the accuracy of the foreground ISM estimations by comparing the structure of the Galactic foreground ISM on half a degree (LAB \HI\ data) and arcminute scales (AKARI FIS all-sky far-infrared data, \citealt{doi2015}). They found that in some cases the well resolved Galactic foreground column densities may be 50\% off the value of the low resolution data. Using 2' resolution AKARI FIS data \cite{Toth2018AKARI} located a local maximum of $N($H$)_\mathrm{MW}$ in the direction of the long GRB source GRB051022. They found that the high resolution estimate of the foreground column density is higher than the value used by the UKSSDC. As a consequence of this, the intrinsic $N($H$)_\mathrm{GRB}$ value became slightly lower, than the former estimate \citep{Toth2018AKARI}. Another example is the case of GRB110213A where the UKSSDC estimate of the intrinsic $N($H$)_\mathrm{GRB}=3.6\times 10^{20}$\,cm$^{-2}$ is unlikely low. But the high resolution AKARI FIS data uncovered  a "hole" in the Galactic ISM, which resulted in a more accurate foreground estimate and a more realistic intrinsic $N($H$)_\mathrm{GRB}$ \citep{Toth2018IAUS}.  
	
These examples indicate the need for an angular resolution of around 1-2 arcmin to be able to estimate correctly the Galactic foreground column density in the line of sight of the GRBs. This resolution can be achieved with the new FIR maps. However, FIR radiation alone is not always a good tracer for the Galactic hydrogen column density. An example for this is the dust-to-gas ratio variations in Galactic clouds at various latitudes (e.g. \citealt{Burstein1978, Paradis2012, chen2015}). The relation between extinction A$_V$ and \HI\ column density $N{\rm (H)}$, is not expected to be very tight, as the dust is likely more closely correlated with the molecular phase H$_2$, in addition \HI\ and H$_2$ are not necessary closely correlated (e.g. \citealt{Boulanger1996, Wolfire2010, Lenz2017}).
	
Motivated by these facts, we performed a pilot study with high spatial resolution \HI\ measurements towards 4 GRBs as a first step towards establishing a reliable foreground correction. In this paper, we present new \HI\ maps from our observations alongside with estimating the optical depth of the gas. We calculate the foreground and the intrinsic hydrogen column density for the GRBs and compare it with calculations based on single dish \HI\ and FIR data.
	
This paper is structured the following way: in section~\ref{sec:Data} we describe our target selection, our observations with the ATCA and the additional \HI\ and infrared data that we use. In section~\ref{sec:Foreground_column_density} we describe and discuss the \HI\ and 1.4 GHz continuum maps and the calculated foreground column densities. In section~\ref{sec:Intrinsic_column_density} we present the newly derived intrinsic hydrogen column densities from the newly fitted X-ray spectra of the GRBs. In section~\ref{sec:Fluctuations} we analyse the fluctuations in the foreground data and in section~\ref{sec:Summary} we summarise our results.
	
\section{Data}
\label{sec:Data}
	
In this section we describe our target selection, the ATCA observations and the archival \HI, infrared and X-ray data that we use in this paper.
	
\subsection{Target selection}
	
For the target selection, we defined a parent sample of the strongest X-ray events observed by the Swift XRT telescope that have spectroscopic redshift measurements and precise X-ray and optical position, with positional errors of a few arcseconds. This selection ensures that the calculated X-ray spectra have the best S/N ratio. From this parent sample we further selected 4 sources, that were observable with the ATCA, for a preliminary survey. Our observed sources are: GRB070508, GRB081008, GRB100425A and GRB100621A. These sources are all long-duration GRBs with redshifts between 0.54 and 1.96. All of them are at high Galactic latitudes and have slightly different structures in their Galactic foreground (see Appendix A for large scale foreground maps, Figs.~\ref{fig:GRB070508_maps}, ~\ref{fig:GRB081008_maps}, ~\ref{fig:GRB100425_maps}, ~\ref{fig:GRB100621_maps}).
	
\subsection{\HI\ data}
	
We obtained high resolution synthesis \HI\ line observations for the sample with the Australia Telescope Compact Array (ATCA). We used a single pointing for each source. Details of the observations are given in Table~\ref{tab:observations}. The ATCA is a radio-interferometer consisting of six 22 m dishes, creating 15 baselines in a single configuration. While five antennas (CA01-CA05) are reconfigurable along a 3 km long east-west track (and a 214 m long north-south spur), one antenna (CA06) is fixed at a distance of 3 km from the end of the track creating the longest baselines. 
	
We used the 1M-0.5k correlator configuration on the Compact Array Broad-band Backend (CABB; \citealt{Wilson2011}) with a 3 MHz wide zoom band\footnote{The 3 MHz wide zoom band consists of 5 concatenated 1 MHz zoom bands, each with 2048 channels overlapped by 50 \% to obtain a flat bandpass.} divided into 6145 channels. This gives a velocity resolution of 0.103 \kms. We used two zoom bands for the \HI\ observations centred at 1417 and at 1420 MHz. The 1417 MHz band is used for bandpass calibration. 
	
	\begin{table*}
		\centering
		\caption{HI observations with ATCA.}
		\label{tab:observations}
		\begin{tabular}{l c c c c c c c  }
			\hline
			GRB name & R.A. & Decl. & l & b & array & integration time & date \\
			& [hh:mm:ss] & [dd:mm:ss] & [$^{\circ}$] & [$^{\circ}$] & & [h]  & \\
			\hline
			GRB070508  & 20:51:12 & -78:23:05 &	314.8597 & -32.4170& H 214 & 7.08 & 25,26 Mar 2017 \\
			GRB070508  & 20:51:12 & -78:23:05 &	314.8597 & -32.4170& 6 C & 8.73 & 9,10 Dec 2017 \\
			GRB081008  & 18:39:50 & -57:25:24 & 338.1202 & -21.1077& 750 C & 1.65 & 18 Dec 2015 \\
			GRB081008  & 18:39:50 & -57:25:24 & 338.1202 & -21.1077& H 168 & 4.19 & 16 Sep 2016 \\
			GRB081008  & 18:39:50 & -57:25:24 & 338.1202 & -21.1077& EW 367 & 8.82 & 13,14 Jan 2017 \\
			GRB100425A & 19:56:47 & -26:25:51 & 014.8064 & -25.2862& 750 C & 1.73 & 18 Dec 2015 \\
			GRB100425A & 19:56:47 & -26:25:51 & 014.8064 & -25.2862& EW 367 & 7.29 & 11,12,14 Jan 2017 \\
			GRB100621A & 21:01:13 & -51:06:22 &	347.4378 & -40.8963& H 214 & 5.8 & 25,26 Mar 2017 \\    
			\hline
		\end{tabular}
	\end{table*}
	
Data reduction was carried out with the {\sc Miriad} software package \citep{Miriad}. We used the standard ATCA primary calibrator PKS 1934-638 for bandpass and amplitude calibration. We used standard ATCA calibrator sources for phase calibration, which we list in Tab.~\ref{tab:observations}. The bandpass calibration is not trivial because all potential bandpass calibrators, including PKS 1934-638, show strong \HI\  absorption near $v_{LSR}=0$ \kms. We used two methods to avoid this absorption line: 
\begin{enumerate}
	\item Frequency-switching, using two zoom bands, one centred at 1420 MHz and second one centred at 1417 MHz, where both bands have the same number of channels. This method assumes that the bandpass is the same in the two neighbouring zoom bands. 
	\item Bandpass calibration with PKS 0023-263, a calibrator close to the Galactic Pole with no Galactic \HI\ absorption. 
\end{enumerate}	
We used the frequency switching method for almost all of our observations since PKS 1934-638 (12.6 Jy) is a much brighter sources compared to PKS 0023-263 (6.7 Jy), and is a better bandpass calibrator.  However, for the first test observations in December 2015 the frequency-switching was not included in the calibration and we used PKS 0023-263.
	
After the calibration the ATCA data were combined with single dish data from GASS\footnote{GASS data: \url{https://www.astro.uni-bonn.de/hisurvey/gass/}}. GASS maps the Galactic \HI\ emission across the whole Southern sky ($\delta \leq 1^{\circ}$), with a spatial resolution of ~16', a velocity resolution of 0.826 \kms\ and an rms brightness temperature noise of 57 mK. 
	
To combine the ATCA and GASS data we binned the ATCA channels to a velocity resolution of  1.03 \kms\ for GRB081008 and GRB100425A, and to 0.51 \kms\ for GRB070508 and GRB100621A. We combined the ATCA and the GASS data before the image deconvolution step. We gridded and scaled the GASS data to match the properties of the ATCA data. Instead of clean, we used the maximum entropy method (MEM, {\sc Miriad} task {\sc maxen}), using the scaled GASS cube as the default to help the convergence. We used the {\sc Miriad} task {\sc restor} to make the MEM output cubes (convolved model plus the residual). The MEM cubes were then corrected for the ATCA primary beam, and converted from the Jy beam$^{-1}$ units to $T_{\mathrm{B}}$ in K scale. From hereafter we refer to these combined data as ATCA+GASS. 

We also constructed data cubes from the ATCA data alone, without separating the continuum and the line data. These data cubes are used to search for \HI\ absorption lines. We excluded all baselines shorter than 150 m, to filter out most of the extended \HI\ emission and only retain the \HI\ absorption against the continuum sources. This is necessary, because we used very compact array configurations for our observations. These configurations are optimal for detecting extended emission in the Galaxy, however, without filtering, the extended emission introduces a large amount of noise that washes out the absorption signal. Ideally all short baselines should be excluded, but using fewer baselines decreases the sensitivity of the data cube.
	
\begin{table}
	\centering
	\caption{HI image properties.}
	\label{tab:observations}
	\begin{tabular}{l c c}
		\hline
		Name & Synthesized beam & Phase calibrator\\
		&  & \\
		\hline
		GRB070508 & 3.64' $\times$ 1.92' & PKS2142-758\\
		GRB081008 & 2.76' $\times$ 2.12' & PKS1814-637\\
		GRB100425a & 3.77' $\times$ 1.49' & PKS0023-263\\
		GRB100621a & 2.86' $\times$ 1.7' & PKS2052-474\\
		\hline
	\end{tabular}
\end{table}
	
In this paper we also use data from the Leiden/Argentine/Bonn (LAB) Survey of Galactic HI \footnote{\HI\ profiles: \url{https://www.astro.uni-bonn.de/hisurvey/profile/}} \footnote{\HI\ maps: \url{https://heasarc.nasa.gov/Tools/w3nh\_maps.html}} \citep{kalberla2005}. The LAB Survey provides \HI\ data for the whole sky with and angular resolution of 0.6$^{\circ}$, a velocity resolution of 1.3 \kms\ and an rms brightness-temperature sensitivity of 70-90 mK. In Appendix~\ref{appendix:large_scale} and in Section~\ref{sec:Foreground_column_density} we show data from the \HI4$\pi$ survey (HI4PI) which is a combination of the GASS survey and the Effelsberg-Bonn \HI\ Survey (EBHIS; \citealt{Winkel2016}). HI4PI covers the whole sky like LAB, but has better spatial resolution (16.2'), velocity resolution (1.29 \kms) and sensitivity (43 mK). Additionally, it also has full angular sampling. 
	
\subsection{Far-infrared data}
	
We have obtained far-infrared data from the Planck Legacy Archive\footnote{\url{https://pla.esac.esa.int/}}. The Planck satellite \citep{Tauber2010} has provided an all-sky sub-millimetre and millimetre survey ranging from 30 GHz to 185 GHz.
	
For our analysis we use $6.5^\mathrm{o} \times 6.5 ^\mathrm{o}$ Planck maps towards the four observed GRBs. There are three different versions of the Planck all-sky dust maps available: 
\begin{inparaenum}
	\item an E(B-V) map from the first data release (PR1;\citealt{planck2014}) and two A$_V$ maps from the second data release, 
	\item A$_V$ normalized to SDSS QSOs - RQ and 
	\item A$_V$ using the DL dust model \citep{Draine2007} (DL; \citealt{planck_dust}).
\end{inparaenum}
	
\subsection{X-ray data}
	
In this work we used X-ray data from the Neil Gehrels Swift Observatory (formally known as 'Swift' space telescope). Swift has 3 instruments on board: the BAT (Burst Alert Telescope a gamma-detector; \citealt{BAT_1, BAT_2}), the XRT (X-Ray Telescope; \citealt{XRT,XRT2}), and the UVOT (Ultraviolet/Optical Telescope; \citealt{UVOT1}). The X-ray telescope can take simple images, light-curves and spectra for the GRBs' afterglow. Normally it is used in the energy range of 0.3-10.0 keV. The X-ray telescope can observe in 3 operating modes depending on the flux. In this work we analyse the Photon Counting (PC) mode data, in this mode fluxes range from the XRT's sensitivity limit of $2 \times 10^{-14}$ to $3 \times 10^{-11}$ erg cm$^{-2}$ s$^{-1}$. In this mode 2 dimensional images are taken with a time resolution of 2.5 seconds and approximately 260 eV energy resolution. The X-ray spectra are created automatically from the event data by HEASARC HEASOFT\footnote{\url{https://heasarc.gsfc.nasa.gov/docs/software/heasoft/}} software \citep{Evans2009}. These spectra can be downloaded from the Swift archive (UK Swift Science Data Centre\footnote{\url{http://www.swift.ac.uk/archive/}}). Usually several types of spectra are available in the catalogue. Almost all sources have time-averaged spectra and if the GRB is observable for more than 4 ks after the trigger, a further late-time spectra is also available. We used the PC mode time averaged spectra for our analysis in this paper. 
	
\section{Foreground column density}
\label{sec:Foreground_column_density}

\subsection{Column density from \HI\ data}

\begin{figure*}
	\centering
	\subfigure{\includegraphics[width=6.2cm]{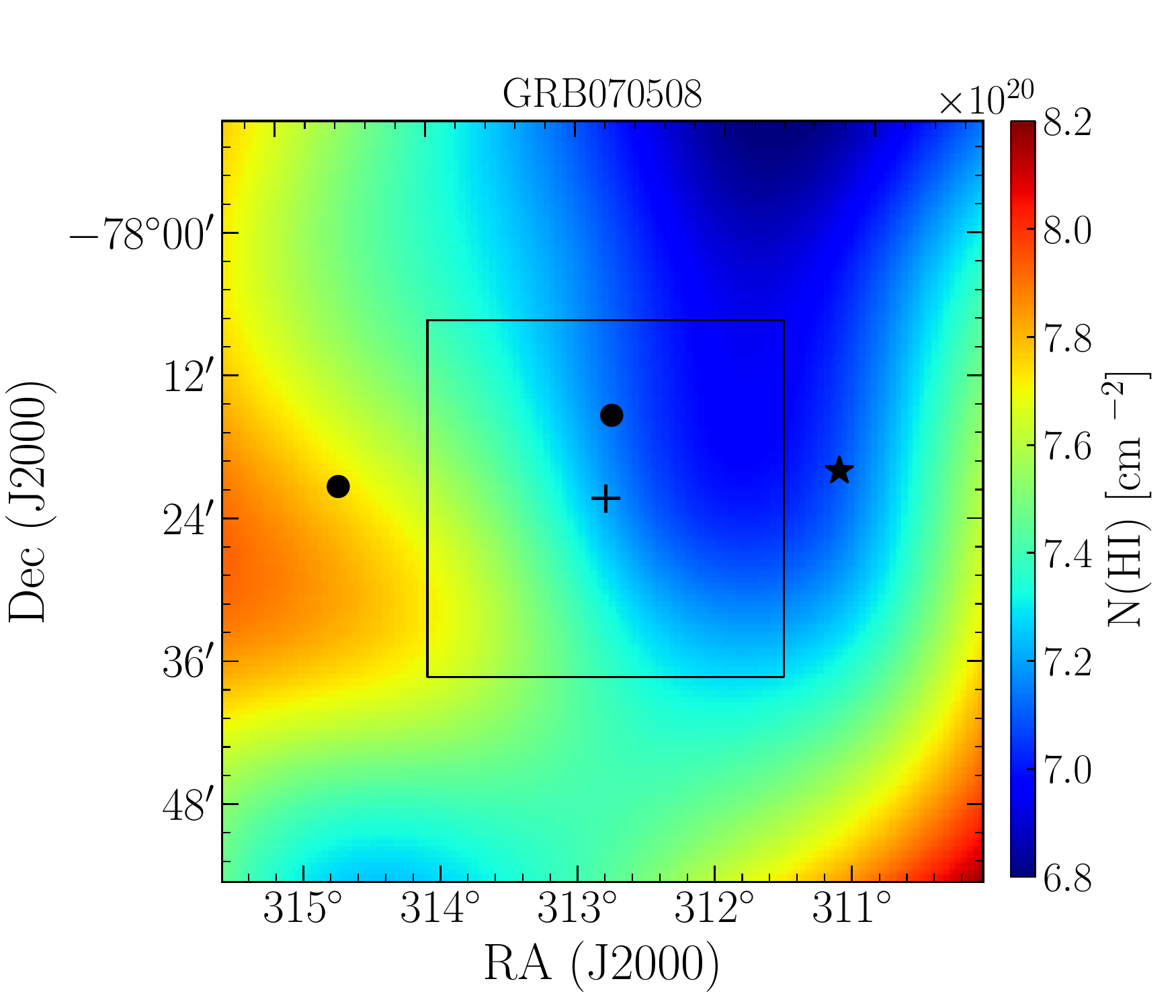}}
	\subfigure{\includegraphics[width=6.2cm]{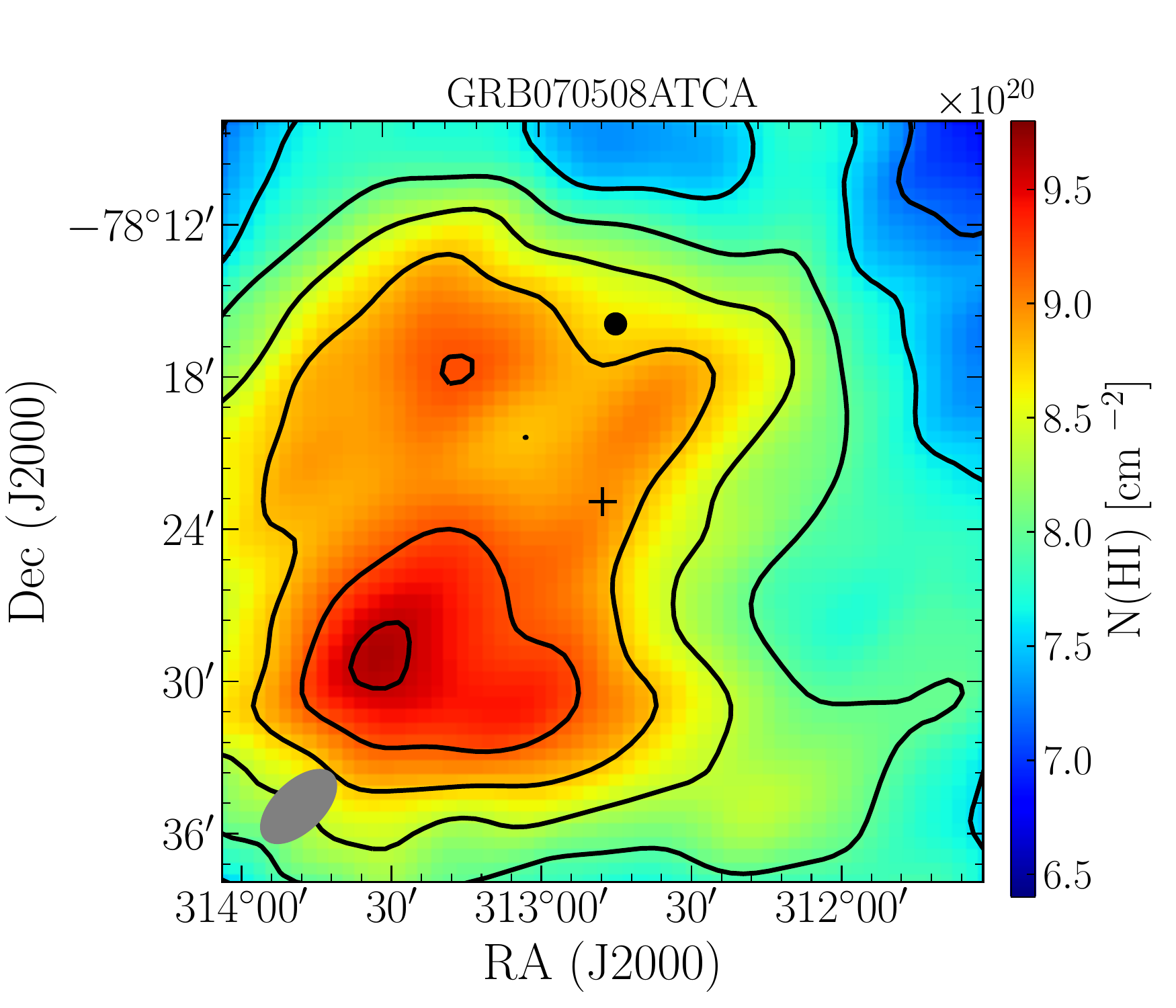}}		
	\subfigure{\includegraphics[width=6.2cm]{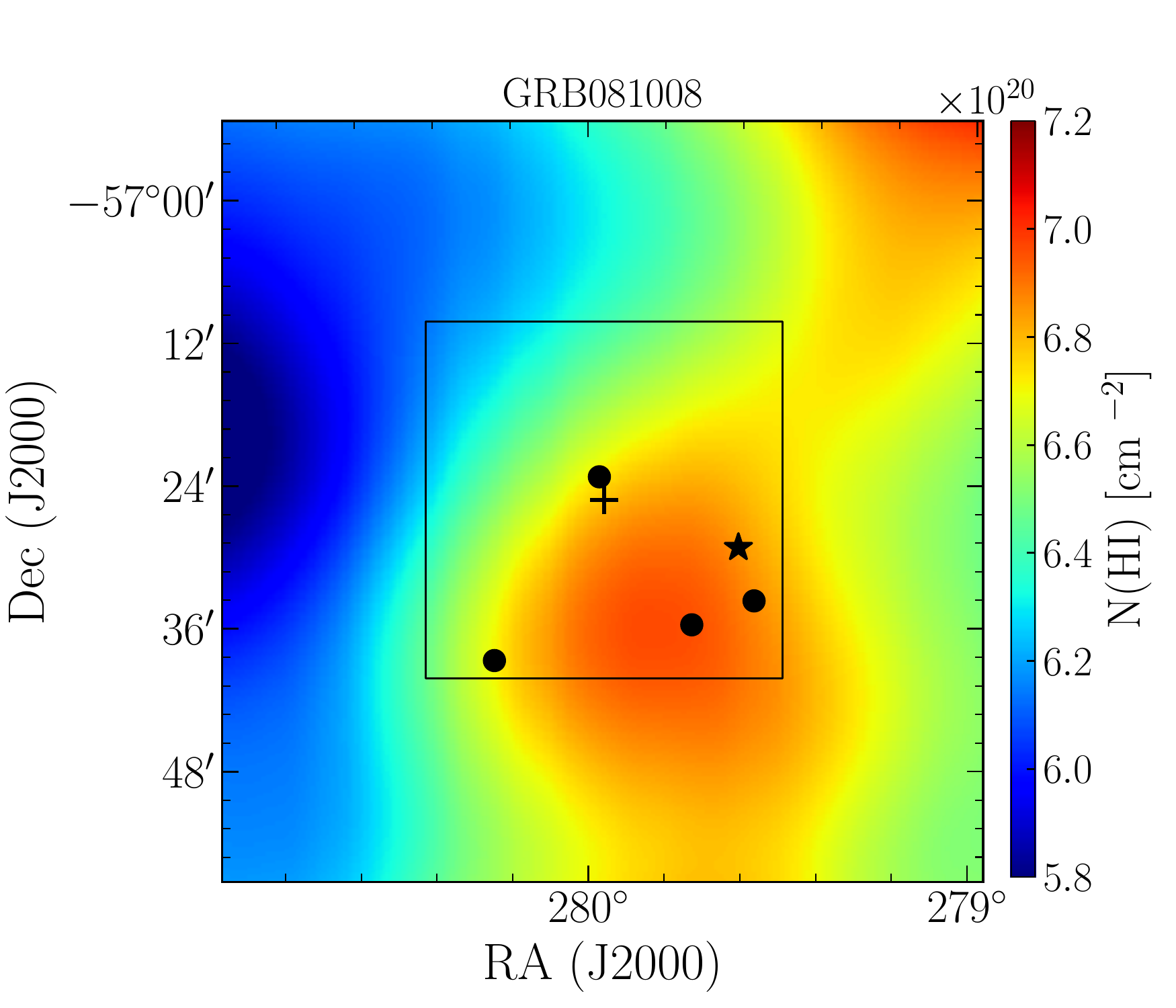}}
	\subfigure{\includegraphics[width=6.2cm]{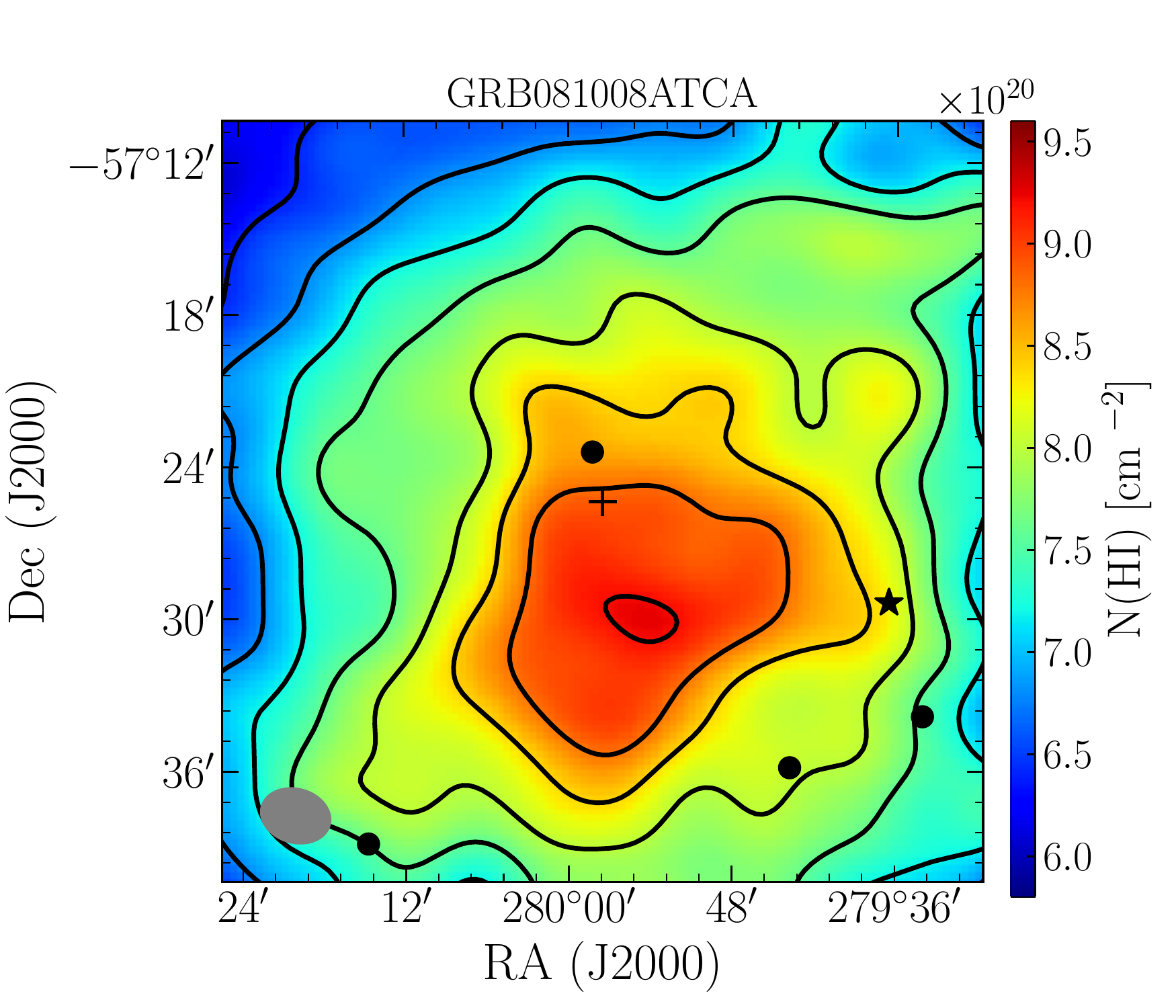}}	
	\subfigure{\includegraphics[width=6.2cm]{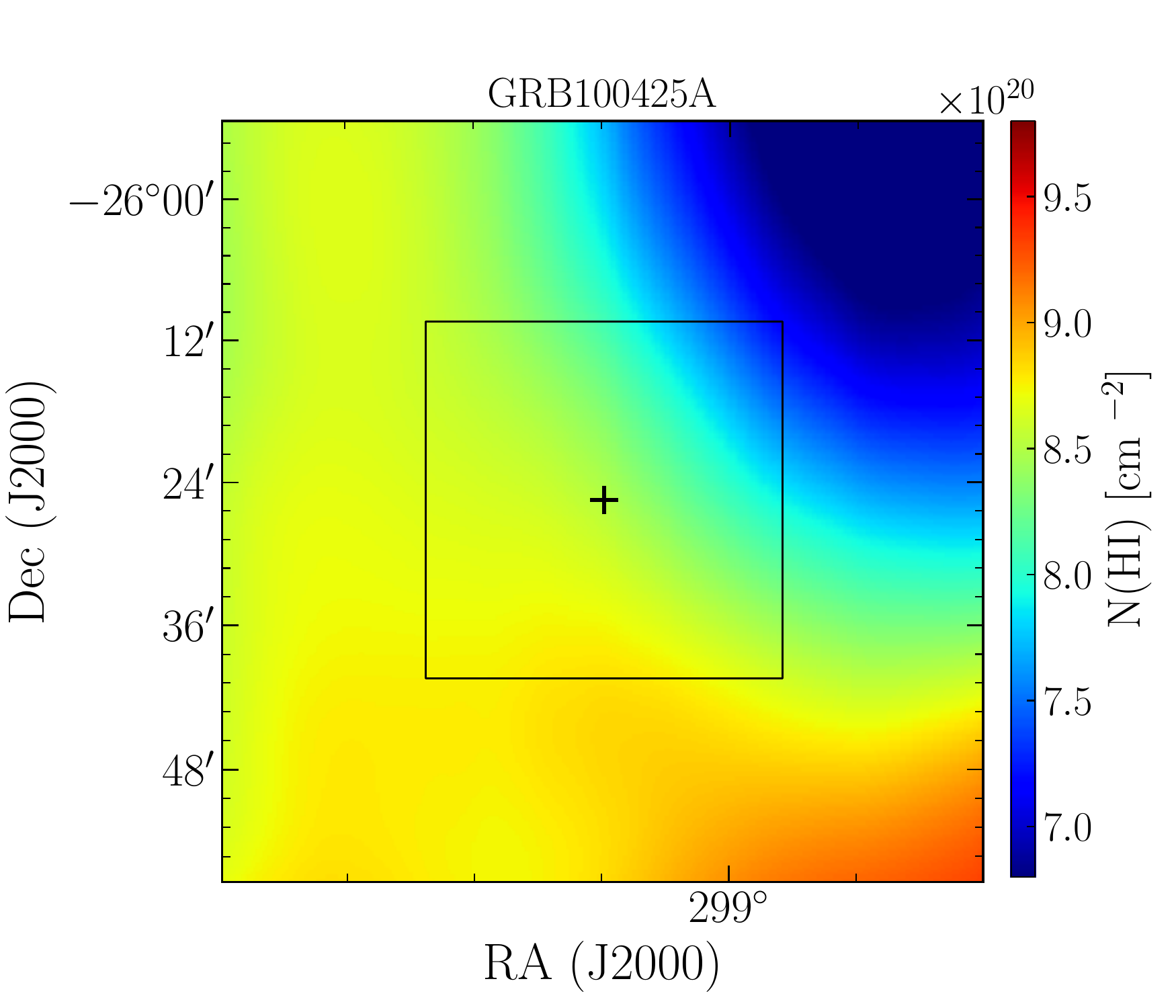}}
	\subfigure{\includegraphics[width=6.2cm]{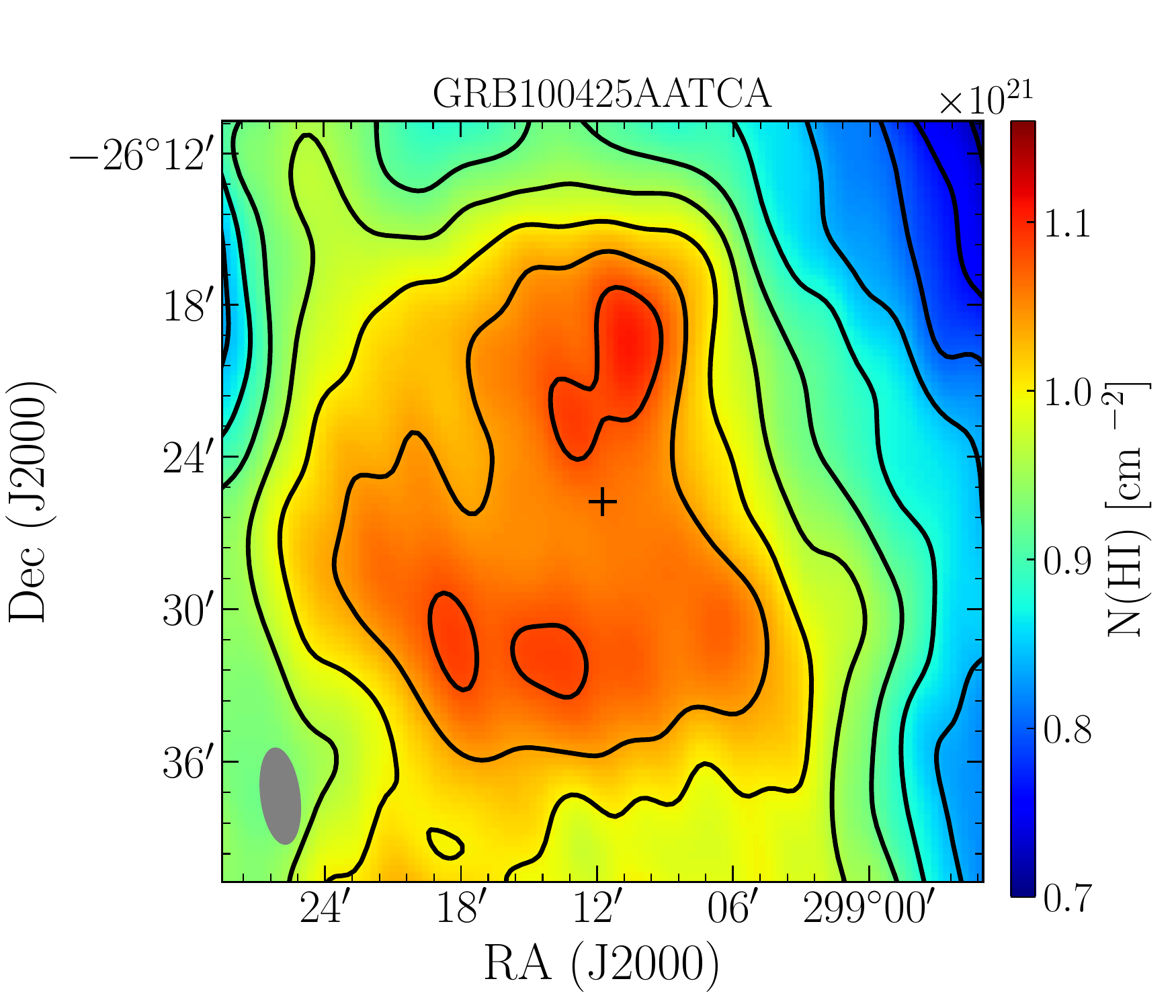}}	
	\subfigure{\includegraphics[width=6.2cm]{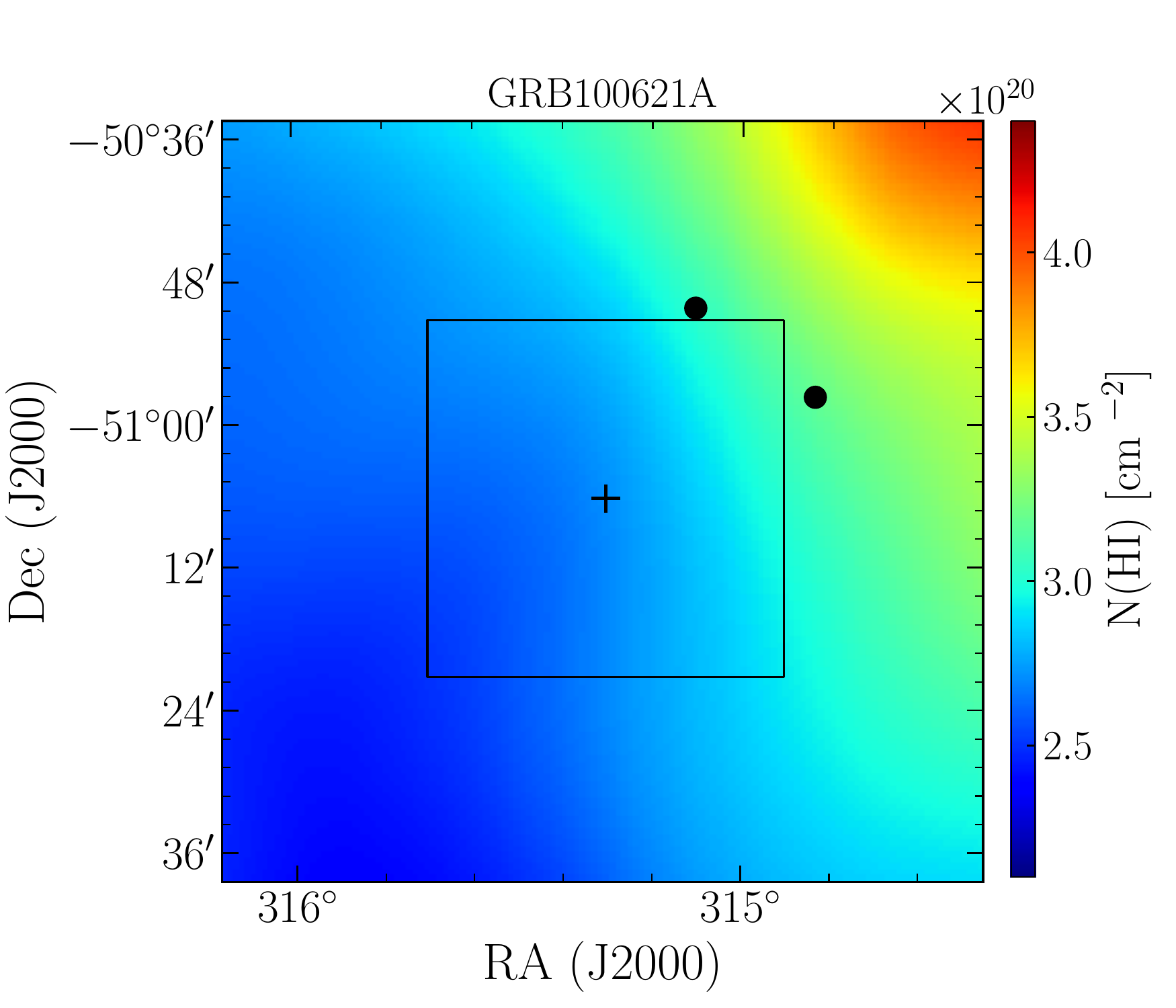}}
	\subfigure{\includegraphics[width=6.2cm]{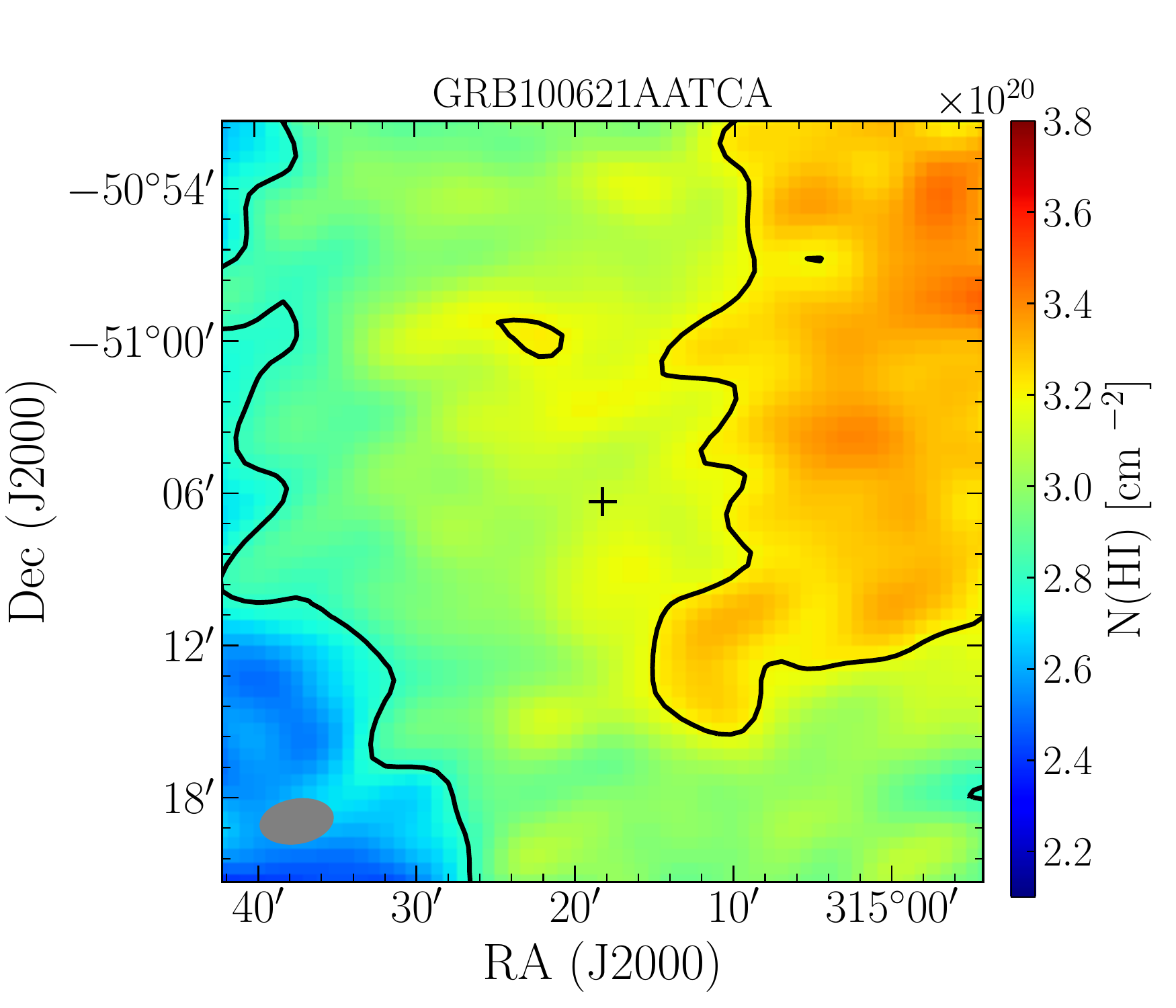}}
	\caption{Column density maps from the GASS (left) and the combined ATCA+GASS (right) data. The black square on the left shows the size of the ATCA image (0.5 deg). A cross marks the position of the GRB in each image, stars mark the positions of the continuum sources with detected absorption lines and circles mark the positions of continuum sources with no detected absorption lines. The grey oval on the ATCA+GASS images indicates the size of the synthesised beam. Contour levels start at $2 \times 10^{20}$ cm$^{-2}$ and increase by $0.4 \times 10^{20}$ cm$^{-2}$ steps.}
	\label{fig:HI_maps}
\end{figure*}

The column density of optically thin ($\tau = 0$) \HI\ gas is calculated the following way:
\begin{equation}
N(\mathrm{HI}) = C_{0} \int T_{B}(v) dv,
\end{equation}
where $C_{0} = 1.823 \times 10^{18}$ cm$^{-2}$K$^{-1}$ (\kms) and $T_{B}$ is the brightness temperature. We derive the foreground \HI\ column densities for each of the ATCA+GASS data cubes. In Fig.~\ref{fig:HI_maps} we present the column density distribution maps of the ATCA+GASS data. Each observed field has \HI\ clouds on the angular scale of a few arc minutes. We detect the brightest \HI\ emission in the field of GRB100425A and the weakest emission in the field of GRB100621A. We also include the extended GASS $N{\rm (HI)}$ maps around the GRB positions to show that the higher resolution maps uncover the structure of the ISM in much more detail. Note that we adjusted the colour scale of each image to emphasise the regions with the bright \HI\ emission.
	 
\subsubsection{Considering optical depth}

When calculating $N{\rm (HI)}$ from \HI\ emission measurements, one of the uncertainties is the optical depth. If the optical depth is not negligible ($\tau > 0$) we need to correct the \HI\ emission for absorption to get the correct column density. The challenge in this is that the optical depth can not be measured from single dish \HI\ emission observations, which is the reason why it is usually assumed to be negligible. However, this assumption is not correct in the Galactic Plane and in the vicinity of molecular clouds (e.g. \citealt{Dickey2003,Bhir2015}). In these regions a significant fraction of the gas can be cold, dense, and optically thick. To directly measure the optical depth of the gas, \HI\ absorption against bright continuum sources can be used. With such an absorption spectra a more accurate \HI\ column density can be calculated. Assuming a one phase ISM the column density is:
\begin{equation}
N(\mathrm{HI})_{\mathrm{corrected}} = C_{0} \int_{\Delta V} T_{B}(\nu) \frac{\tau}{1-e^{-\tau}} d\nu,
\label{eq:NHI_corrected}
\end{equation} 
where $\tau$ is the optical depth and $\Delta V$ is the range of integration.
	
\begin{figure*}
		\centering
		\subfigure{\includegraphics[width=8.4cm]{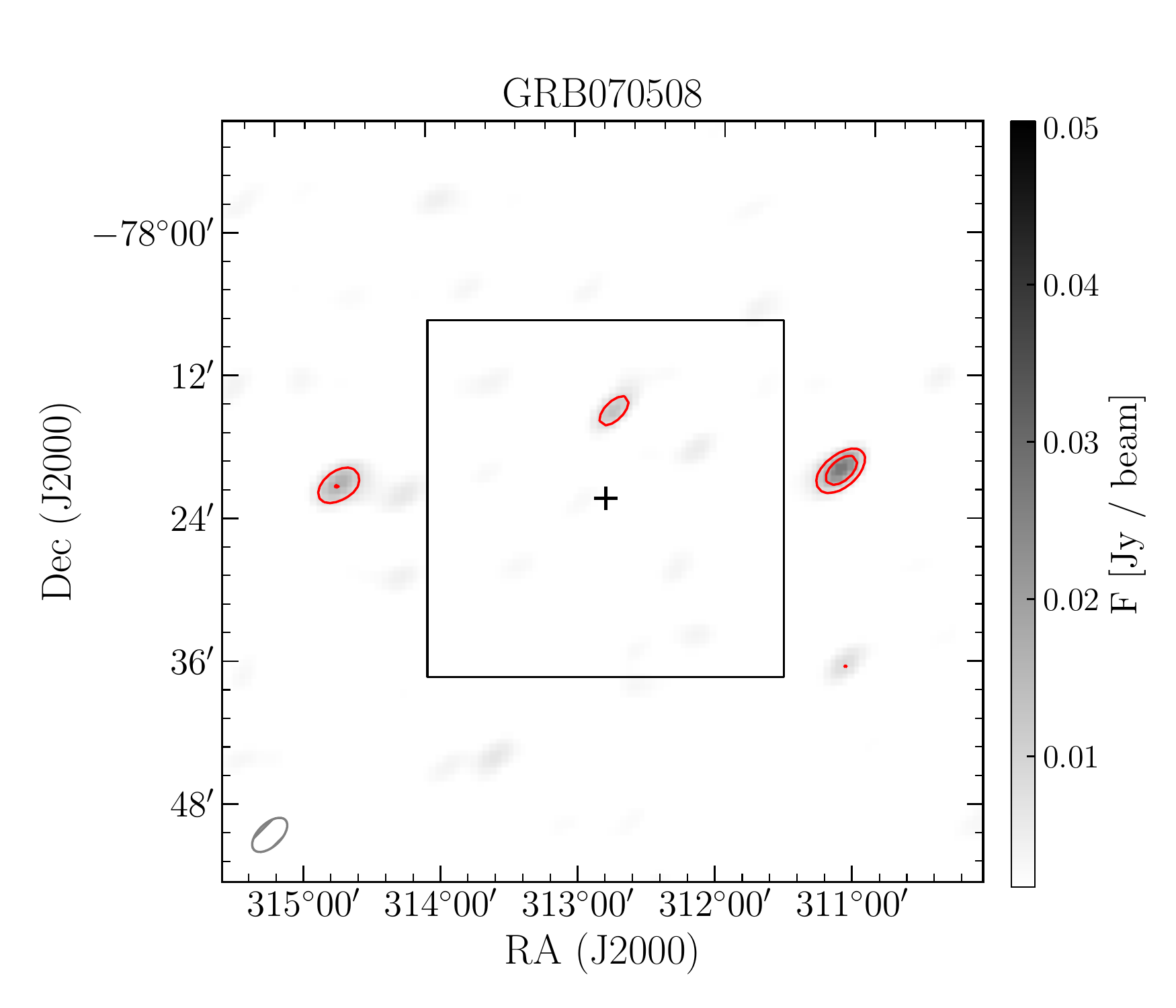}}
		\hfill
		\subfigure{\includegraphics[width=8.4cm]{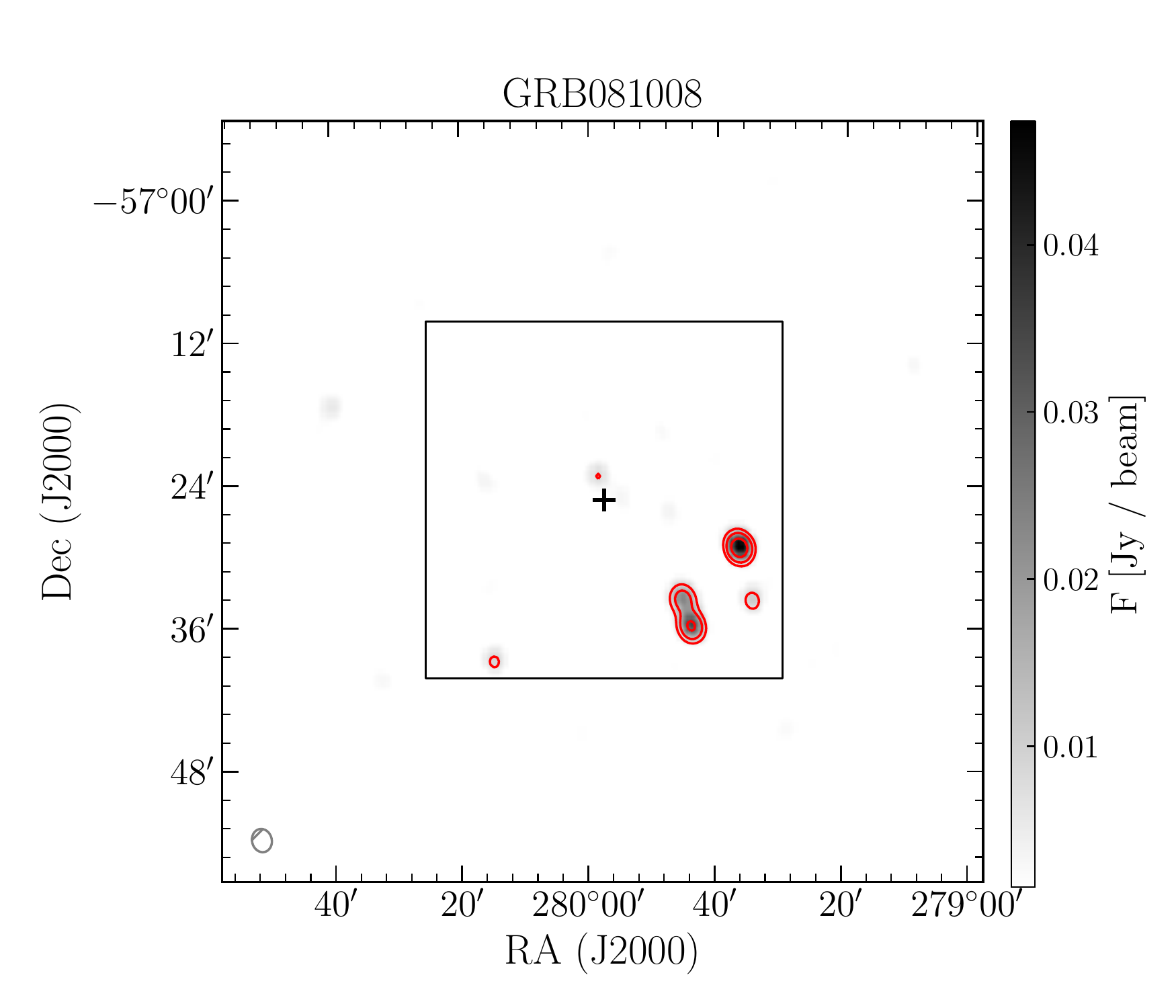}}
		\hfill
		\subfigure{\includegraphics[width=8.4cm]{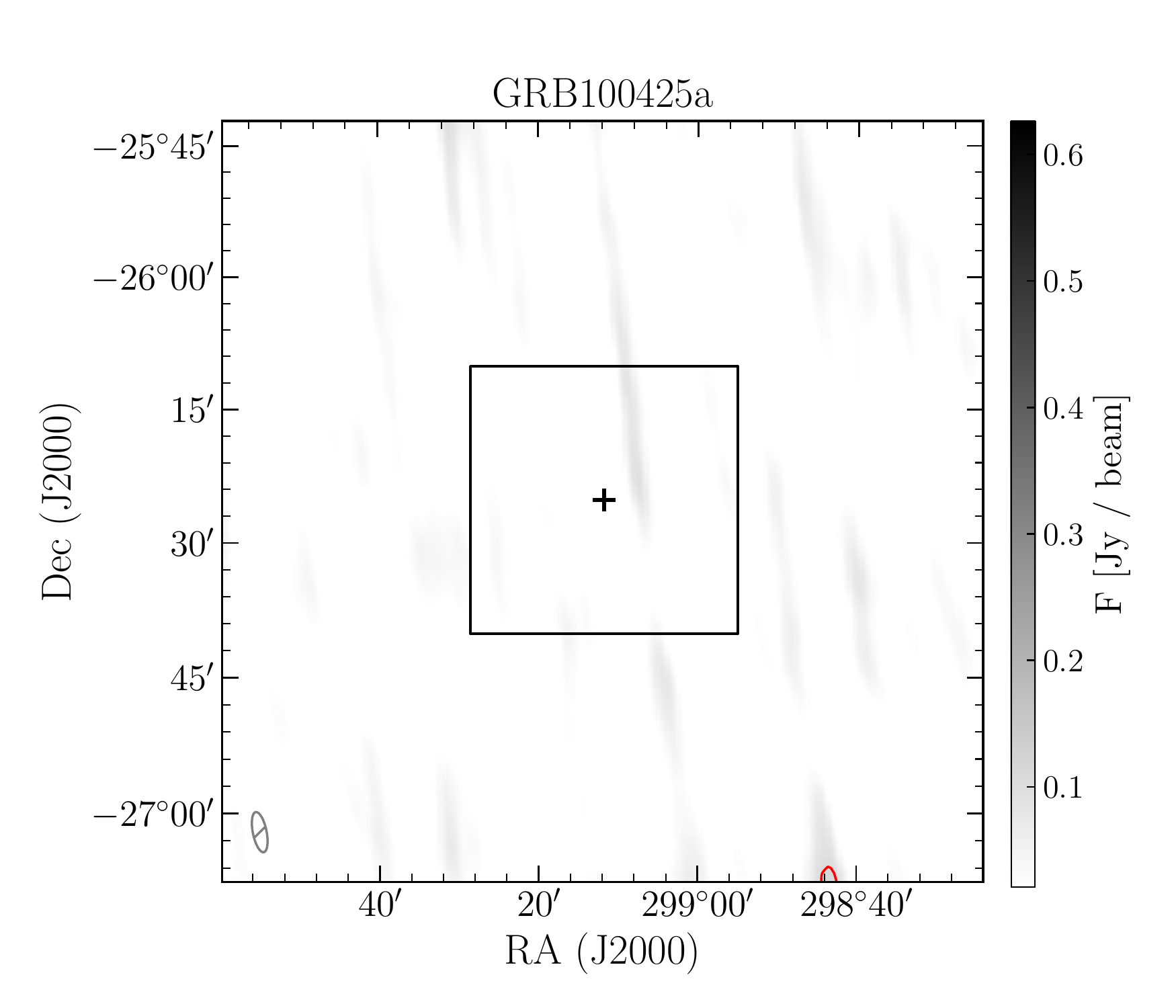}}
		\hfill
		\subfigure{\includegraphics[width=8.4cm]{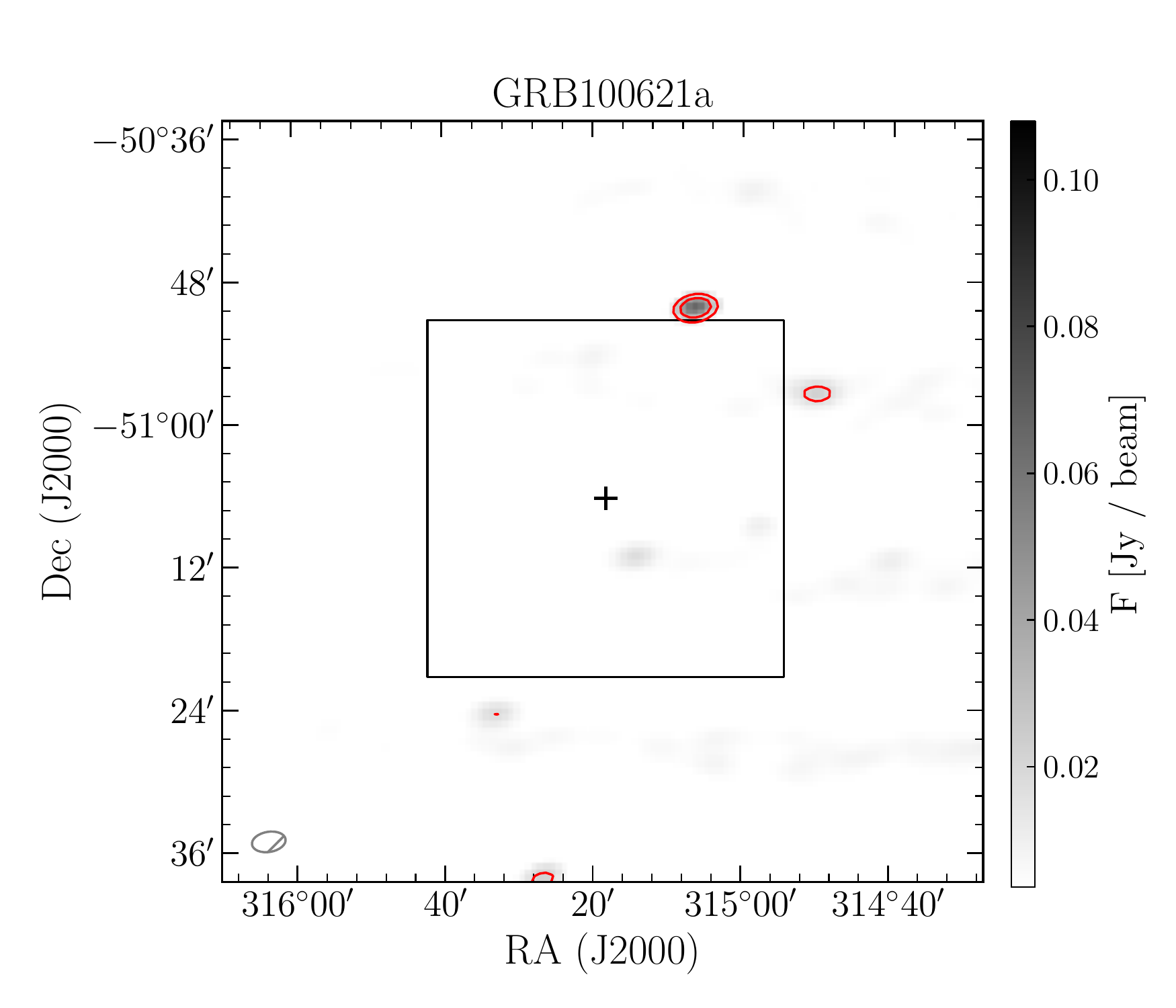}}
		\caption{1.4 GHz continuum maps. Contour levels are $5\sigma, 10\sigma, 20\sigma$. The grey ellipses in the bottom left corner show the size of the synthesized beam. The black squares mark the 0.5$^{\circ}$ field shown in Fig.~\ref{fig:HI_maps}. We use the brightest continuum source in the field to estimate the \HI\ optical depth towards GRB070508 and GRB081008. We note that the image around GRB100425A only shows artefacts and noise. This field has the lowest sensitivity of the ATCA observations and we do not detect any continuum sources.}
		\label{fig:continuum}
	\end{figure*}

To investigate if there are bright continuum sources suitable for absorption measurements, we created 1.4 GHz continuum maps from the line free channels of the ATCA data (Fig.~\ref{fig:continuum}). For three of the fields we achieved an rms level on the order of a few mJy. As a result we detect ten continuum sources above 5$\sigma$ in these three fields. We do not detect any significant continuum emission at the position of the GRBs, which could be associated with the GRB host galaxies. Table~\ref{tab:1.4GHz_flux} shows the measured 1.4 GHz continuum fluxes of the detected sources and the 3$\sigma$ upper limits calculated at the positions of the GRBs. We note that in the field around GRB100425A we only see noise in the continuum image. The observations in this field have the lowest sensitivity (rms = 20.88 mJy/beam) and we do not detect any continuum sources.
	
\begin{figure*}
		\centering
		\subfigure{\includegraphics[width=8.4cm]{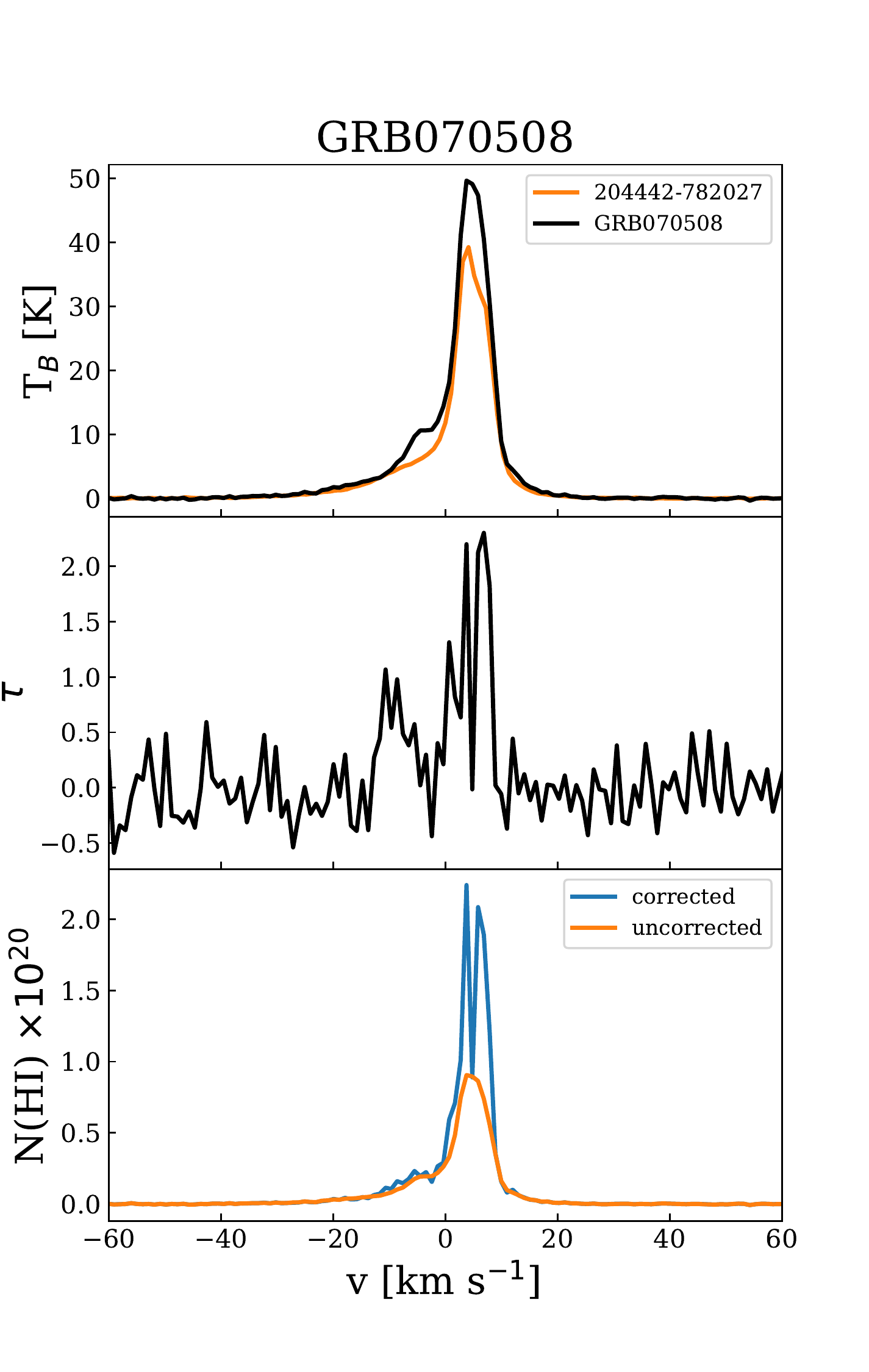}}
		\hfill
		\subfigure{\includegraphics[width=8.4cm]{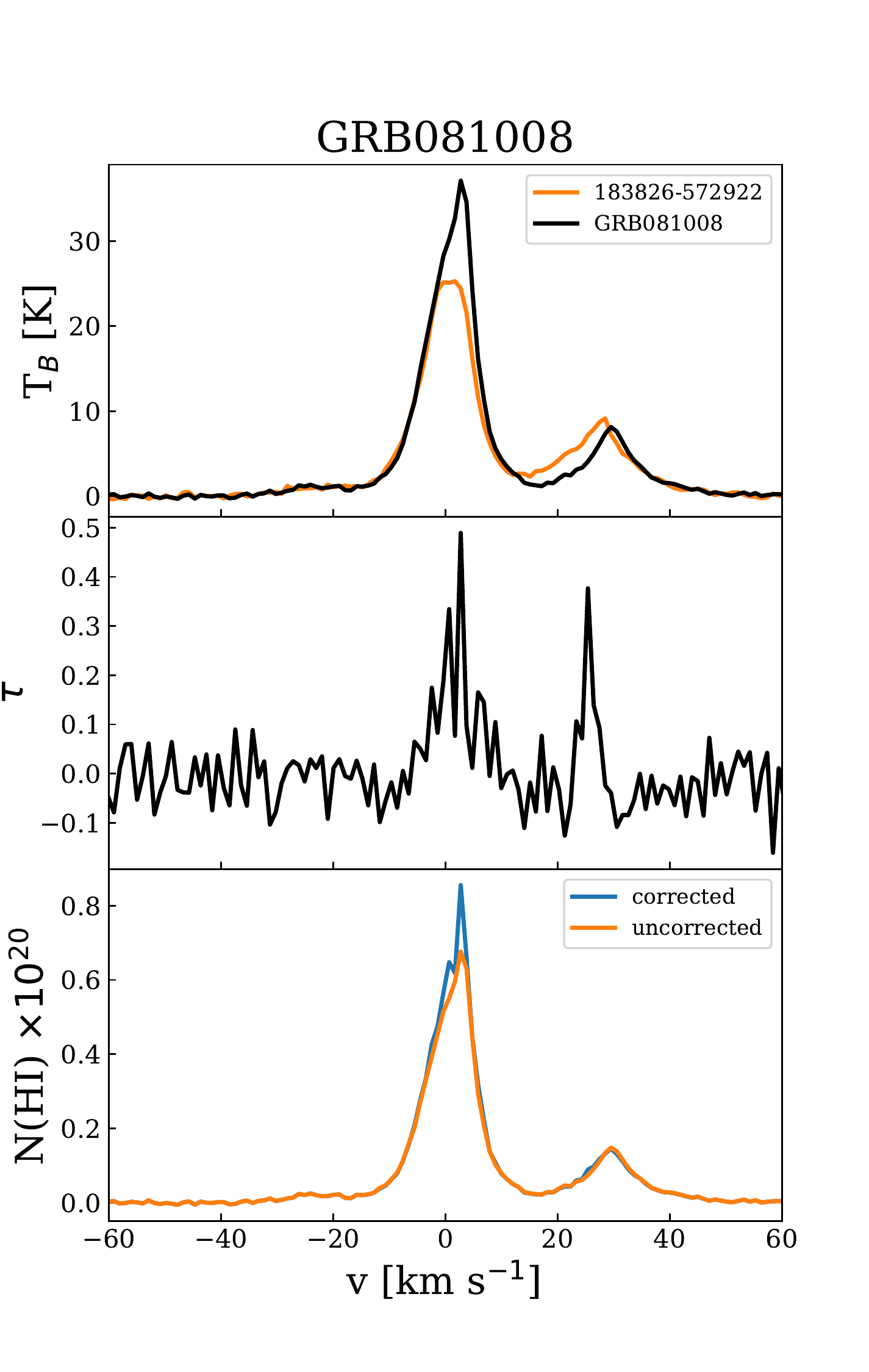}}
		\hfill
		\subfigure{\includegraphics[width=8.4cm]{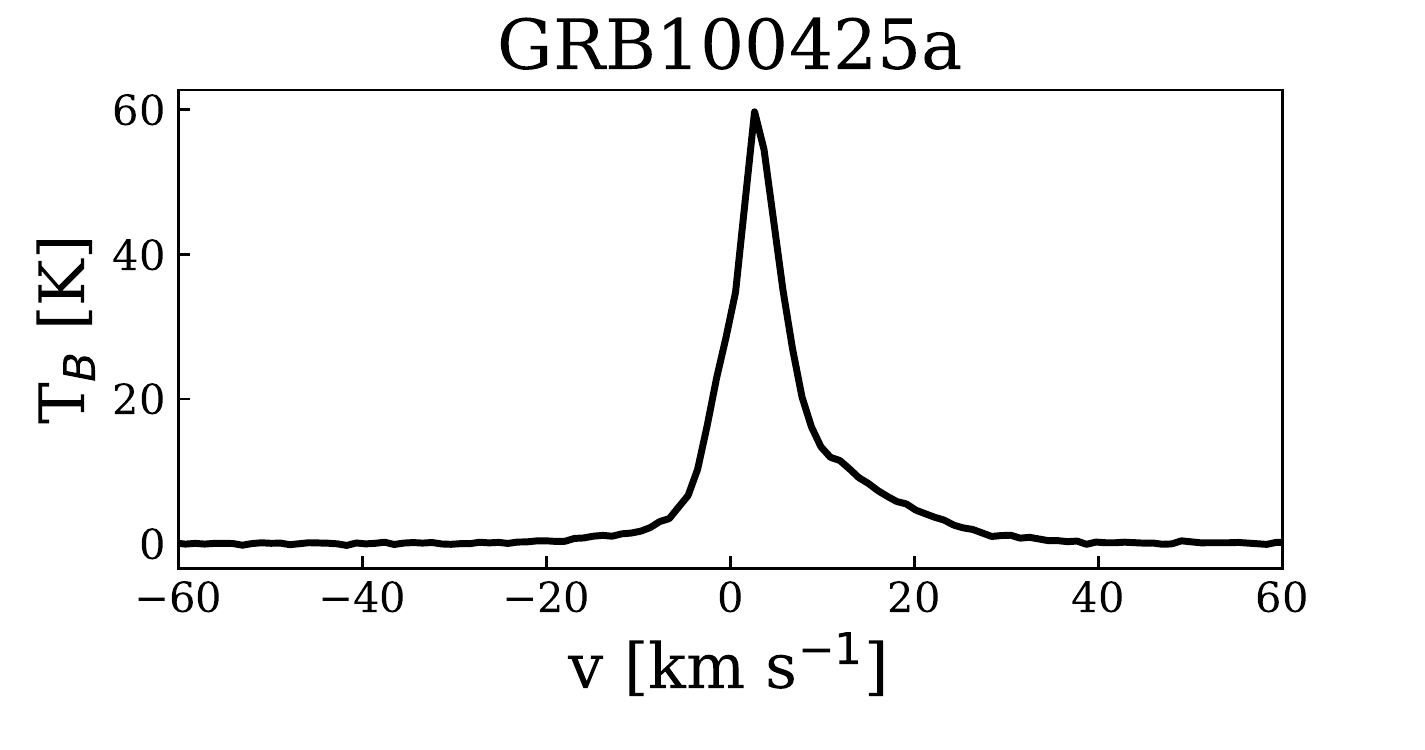}}
		\hfill
		\subfigure{\includegraphics[width=8.4cm]{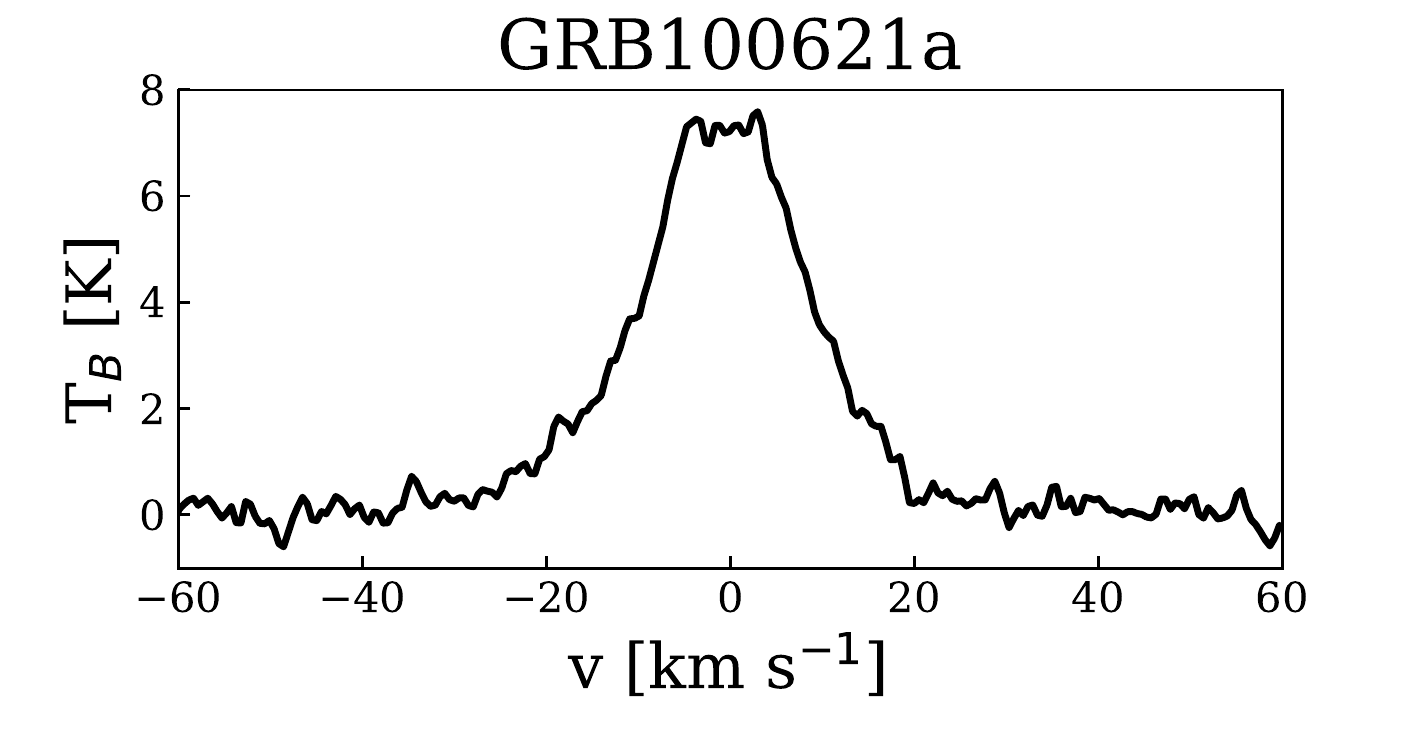}}
		\caption{Brightness temperature, optical depth, and corrected (blue) and uncorrected (orange) $N{\rm (HI)}$ spectra. For GRB100621A and GRB100425A we only show the brightness temperature spectrum.}
		\label{fig:HI_spectrum}
	\end{figure*}

We detect \HI\ absorption lines in two fields around GRB070508 and GRB081008 (Fig.~\ref{fig:HI_spectrum}). In the other two fields the combination of the sensitivity of our observations and the brightness of continuum sources did not result in a line detection. The optical depth sensitivity depends on the continuum flux $(S)$ and the continuum flux sensitivity ($\sigma_{\rm S}$) of the data the following way: $\sigma_{\rm \tau} = \sigma_{\rm S}/S $. Considering that the field around GRB100425A contains the brightest \HI\ clumps in our sample, with a peak intensity of 60 K at the position of the GRB and is also located relatively close to the Aquila-Sagitarius molecular cloud complex, where an extensive amount of dark neutral medium (DNM) was detected by \cite{Grenier2005}, we may expect some optical depth in this field. However, we have the lowest continuum sensitivity in this field and do not detect any continuum sources. In the case of the field around GRB100621A we detect a continuum source, but this field has the lowest \HI\ brightness, with a peak brightness temperature of 8 K, which suggests that the gas is optically thin with no detectable absorption lines. In Tab.~\ref{tab:1.4GHz_flux} we present the optical depth sensitivity for all detected continuum sources and the measured peak optical depth ($\tau_{\rm max}$) for the two sources where we detected \HI\ absorption (J204442-782027, J183826-572922). We also derive 5$\sigma$ upper limits for the sources, with no detected absorption. 

Additionally, Tab.~\ref{tab:1.4GHz_flux} contains the peak brightness temperature ($T_{B,max}$) at the position of each source and the calculated spin temperature ($T_s$). We estimate the line of sight harmonic mean spin temperature the following way:

\begin{equation}
T_{s} = \frac{\int T_{B}(v) dv}{\int (1-e^{-\tau(v)})dv },
\end{equation}

where $T_{B}$ is the brightness temperature of the \HI\ emission and $\tau$ is the optical depth. We calculate $T_{s}$ in the line of sight (LOS) of the two sources with detected absorption lines. We integrate $T_{B}$ and $\tau$ between -20 and 20 \kms, which is the velocity range of the main \HI\ emission feature for these fields. The spin temperatures for the two LOS are 52$\pm$8 K and 367$\pm$40 K, from which the former is a typical temperature for the cold neutral medium (CNM). 
	
\begin{table*}
	\centering
	\caption{Continuum sources detected at 1.4 GHz ($> 6\sigma$), and upper limits for the position of the GRB host galaxies. Columns: (1) field; (2) point source name; (3-4) equatorial coordinates; (5) peak fluxes $S_{peak}$; (6) integrated fluxes $S_{Int}$; (7) optical depth sensitivity ($\sigma_{\rm \tau}$); (8) peak optical depth; (9) brightness temperature $T_B$; (10) spin temperature $T_s$ (integrated between -20 and 20 \kms). 3$\sigma$ upper limits based on the rms of the ATCA continuum images are given for $S_{peak}$ and $\tau_{\rm max}$.}
	\label{tab:1.4GHz_flux}
	\begin{tabular}{l l c c c c c c c c}
		\hline
		Field & Name  & R.A. & Decl. &$S_{\rm peak}$ & $S_{\rm Int}$ & $\sigma_{\rm \tau}$ & $\tau_{\rm max}$ & $T_{B,{\rm max}}$ & $T_{s}$\\
		&&  [hh:mm:ss] & [dd:mm:ss] &[mJy/Beam] & [mJy]& && [K] & [K]\\
		\hline
		GRB070508 & GRB070508  &  &  & $<$ 5.04 &&&& 49.8 $\pm$ 0.1 & \\
		& J204442-782027  &20:44:42 &-78:20:27 & 27.9 $\pm$ 1.7 & 28.72 & 0.06 & 2.3 $\pm$ 0.5 & 45.6 $\pm$ 0.3 & 52 $\pm$ 8\\
		& J205101-781604  &20:51:01 &-78:16:04 & 13.4 $\pm$ 1.7 & 15.85 &0.12& $<$ 0.6 & 49.8 $\pm$ 0.2& \\
		& J205837-782142  &20:58:37 &-78:21:42 & 16.5 $\pm$ 1.7 & 19.57 & 0.1 & $<$ 0.5 & 38.6 $\pm$ 0.4 & \\
			
		GRB081008 & GRB081008  & & & $<$ 4.74 &&&& 37.1 $\pm$ 0.1 & \\
		& J183816-573351  &18:38:16 &-57:33:51 & 10.9 $\pm$ 1.6 & 10.28 & 0.14 & $<$ 0.7 & 23.7 $\pm$ 0.2& \\ 
		& J183826-572922  &18:38:26 &-57:29:22 & 49.9 $\pm$ 1.6 & 49.13 & 0.03 &0.5 $\pm$ 0.1& 25.3 $\pm$ 0.2 & 367 $\pm$ 40\\
		& J183855-573553  &18:38:55 &-57:35:53 & 30.4 $\pm$ 1.6 & 63.59 & 0.05 & $<$ 0.25 & 26.7 $\pm$ 0.2 & \\
		& J183953-572325  &18:39:53 &-57.23.25 & 7.5 $\pm$ 1.6 & 8.2 & 0.19 & $<$ 0.9 & 35.18 $\pm$ 0.2 & \\
		& J184059-573853  &18:40:59 &-57:38:53 & 9.6 $\pm$ 1.6 & 9.0 & 0.17 & $<$ 0.8 & 29.87 $\pm$ 0.2 & \\
			
		GRB100425A & GRB100425A   & & & $<$ 62.64 &&&&  59.69 $\pm$ 0.1 &\\ 
			
		GRB100621A & GRB100621A   & & & $<$ 10.8 &&&& 7.6 $\pm$ 0.2 &\\
		& J205921-505748   &20:59:21 &-50:57:48 & 19.28  $\pm$ 3.6 & 22.96 & 0.19 & $<$ 0.95 & 10.1 $\pm$ 0.8&\\   
		& J210025-505021   &21:00:25 &-50:50:21 & 71.16  $\pm$ 3.6 & 78.70 & 0.05 & $<$ 0.25 & 8.2 $\pm$ 0.2&\\
		\hline
	\end{tabular}
\end{table*}
	
We calculate $N{\rm (HI)}_{\rm ATCA+GASS}$ for all GRB line of sights assuming $\tau = 0$ (Tab.~\ref{tab:columndensity}). We note that, the ATCA values are 1.2 times higher compared to the HI4PI values for all sources, which is due to the scaling of the data when combining interferometric and single measurements. For GRB070508 and GRB081008 we also calculate $N{\rm (HI)}_{\rm corrected}$ using $\tau$ derived from the \HI\ absorption against nearby continuum sources (Tab.~\ref{tab:columndensity}). For the optical depth correction we make the approximation that the optical depth does not change across $\sim 30$ arc seconds. This is necessary since there are no strong detected continuum sources at the positions of the GRBs. We would like to point out that this is a rough approximation since significant optical depth variations ($\Delta \tau$=0.1-0.5) have been observed before on scales of a few arc seconds or even on subarcseconds (e.g. \citealt{Davis1996,Stanimirovic2018}) in certain regions of the sky. Nevertheless, having an approximation of the optical depth, can still improve the foreground column density estimation. In Fig.~\ref{fig:HI_spectrum} we compare the \HI\ emission spectrum (ATCA+GASS) at the position of the GRB with the spectrum at the position of the strongest continuum source. The spectra are similar in shape and intensity, with a $\sim$ 10 K difference in $T_{B}$.  
	
Fig.~\ref{fig:HI_spectrum} also shows the optical depth (ATCA) and the calculated column density spectrum for GRB070508 and GRB081008. For GRB100425A and GRB100621A we only show the \HI\ emission spectrum. The optical depth spectrum shows at least 2 distinct cold components for both detections, which correspond to peaks in the \HI\ emission spectrum. In terms of the $N{\rm (HI)}$ the correction is relatively low for GRB081008 with a $\sim$5\% increase, but it is quiet substantial for GRB070508 with a $\sim$60\% increase (Tab.~\ref{tab:columndensity}). The LOS of GRB070508 is relatively close to the outskirts of the Chamaeleon molecular cloud complex. This molecular cloud complex has a large envelop of DNM \citep{Grenier2005, Grenier2015}, and is associated with an extended cold region in far-infrared \citep{toth2000}. This strongly suggests the presence of optically thick \HI\ in this region, which is also supported by the high optical depth measured by our observations.  
	
To compare the Galactic foreground column densities derived from \HI\ and infrared data, we assume that $N{\rm (HI)} = N{\rm (H)}$ in the regions of our observations. Considering that there are no molecular clouds directly overlapping with our lines of sight we can assume that there is no significant H$_2$ present in these regions. We also make this assumption for GRB070508 considering that the carbon-monoxide (CO) detections in the Chamaeleon complex are several degrees away and the nature of the DNM is still not entirely understood. 
	
\subsection{Foreground column density from Planck data} 
	
We calculated $N(\mathrm{H})$ from the Planck 1st Public Release $E(\mathrm{B}-\mathrm{V})$ reddening values \citep{planck2014}; the 2nd public Release visual V band extinctions $A_{\rm V}({\rm DL})$; and from the $A_{\rm V}({\rm RQ})$ renormalized extinctions, that matches extinction estimates for quasars derived from the Sloan Digital Sky Survey data \citep{planck_dust}. The Planck maps are available at 5 arc minutes angular resolution.
	
We derived the hydrogen column densities from the reddening and extinction values the following way \citep{Guver2009}:
\begin{equation}
N(\mathrm{H})_{\rm Planck} = 2.21 \times 10^{21}A_{\rm V}
\end{equation} 
and
\begin{equation}
N(\mathrm{H})_{\rm Planck} =  6.86 \times 10^{21} E(\mathrm{B}-\mathrm{V}),
\end{equation}
where $A_{\rm V}$ is the $V$-band extinction and $E(\mathrm{B}-\mathrm{V})$ is the reddening.
	
In Tab.~\ref{tab:columndensity} we compare the derived column densities from the ATCA+GASS, HI4PI, LAB, and Planck data. We find that the Planck $N{\rm (H)}_{PR1}$ and $N{\rm (H)}_{\rm RQ}$ agree the best with the column density derived from \HI\ data. This is also true for the spatial distribution of the gas (Fig.~\ref{fig:GRB070508_correlation},~\ref{fig:GRB081008_correlation},~\ref{fig:GRB100425_correlation},~\ref{fig:GRB100621_correlation}). We also find that $N{\rm (H)}_{\rm DL}$ is by a factor of 2 higher compared to all other column densities for 3 of the GRBs. This factor of two difference was also described in \cite{planck_dust}, where $A_{\rm V}$ calibrated with the DL dust model was found to be systematically higher compared to $A_{\rm V}$ derived from optical QSOs in SDSS. \cite{planck_dust} discuss that the reason for the difference is due to the fact that the wavelength of the SED peak seems to trace variations in the FIR opacity of the dust grains per unit $A_{\rm V}$. They rescale $A_{\rm V}$ with an empirical relationship based on optical QSO data to $A_{\rm V}({\rm RQ})$. In agreement with \cite{planck_dust}, we also find that the scaled $A_{\rm V}({\rm RQ})$ data gives better matching results to the \HI\ data. 
   
Overall, the \HI\ and the FIR data give similar results, except for $N{\rm (H)}_{\rm DL}$. The uncertainties of the column densities derived from the Planck data are larger compared to the \HI\ data and $N{\rm (H)}_{\rm DL}$ has the largest uncertainty. Due to our small sample size of four lines of sight we are not able to conclude how well this agreement between the FIR and \HI\ data would hold up for the whole sky. Based on our results we recommend to use the HI4PI data for calculating Galactic foreground column densities for GRBs. In addition, we advise not to use the Planck DL data for Galactic hydrogen column density estimations.  
	
\begin{table*}
	\centering
	\caption{Estimates of the Galactic foreground: reddening and \HI\ column densities from various data. Columns: (1) Name; (2) LAB; (3) HI4PI; (4) the ATCA data not corrected for optical depth; (5) ATCA corrected for optical depth; (6-8) hydrogen column densities based on the Planck survey data.}
	\label{tab:columndensity}
	\begin{tabular}{l c c c c c c c}
	\hline
	Name &$N{\rm (HI)}_{\rm LAB}$ & $N{\rm (HI)}_{\rm HI4PI}$ & $N{\rm (HI)}_{\rm ATCA+GASS}$ & $N{\rm (HI)}_{\rm corrected}$ &  $N{\rm (H)}_{\rm PR1}$ &  $N{\rm (H)}_{\rm RQ}$ & $N{\rm (H)}_{\rm DL}$ \\
	&  [10$^{20}$ cm$^{-2}$] &10$^{20}$ cm$^{-2}$] & [10$^{20}$ cm$^{-2}$] & [10$^{20}$ cm$^{-2}$] & [10$^{20}$ cm$^{-2}$] & [10$^{20}$ cm$^{-2}$] & [10$^{20}$ cm$^{-2}$] \\
	\hline
	GRB070508 & 7.35 $\pm$ 0.09 & 7.108 $\pm$ 0.06 & 9.0 $\pm$ 0.2 & 14.6 $\pm$ 0.2 & 8.16 $\pm$ 0.35 & 10.29 $\pm$ 0.58 & 22.32 $\pm$ 1.34\\
	GRB081008 & 7.63 $\pm$ 0.09 & 6.789 $\pm$ 0.06 & 8.9 $\pm$ 0.1 & 9.4 $\pm$ 0.1 & 7.94 $\pm$ 0.14 & 7.20 $\pm$ 0.76 & 12.80 $\pm$ 1.34\\
	GRB100425A & 8.49 $\pm$ 0.09 & 8.487 $\pm$ 0.06 & 10.6 $\pm$ 0.2 && 10.62 $\pm$ 0.59& 11.64 $\pm$ 0.64 & 22.78 $\pm$ 1.29\\
	GRB100621A & 2.60 $\pm$ 0.09 & 2.695 $\pm$ 0.06 & 3.1 $\pm$ 0.1 && 2.68 $\pm$ 0.95 & 2.41 $\pm$ 0.27 & 4.60 $\pm$ 0.53\\
	\hline
	\end{tabular}
\end{table*}

\section{GRB intrinsic column density from X-ray spectra fitting}
\label{sec:Intrinsic_column_density}

For the fitting of the X-ray spectra we used the {\sc xpec v12.9.1} software\footnote{Xspec is part of the HEASOFT Software package of NASA's HEASARC} which is a widely-used interactive X-ray spectra-fitting program. {\sc xpec} allows to set many physical parameters from cosmology, solar abundances, redshift to ISM absorption models. The UK Swift Science Data Centre uses this software to automatically analyse the Swift XRT spectra of GRBs \citep{Evans2009}. In the automated analysis the used initial settings are the standard flat universe cosmology, low-metallicity ISM abundance and Galactic foreground from the Leiden/Argentine/Bonn (LAB) Survey \citep{kalberla2005}. {\sc xpec} fits a complex theoretical spectrum model that contains a standard powerlaw, flux-convolution model and two ISM absorption models. One of the ISM models is a fixed Galactic component based on the LAB data and the other is a fittable model for the intrinsic column density of the host galaxy.
	
\begin{figure*}
	\centering
	\subfigure{\includegraphics[width=8.4cm, trim= 0 0 100 0]{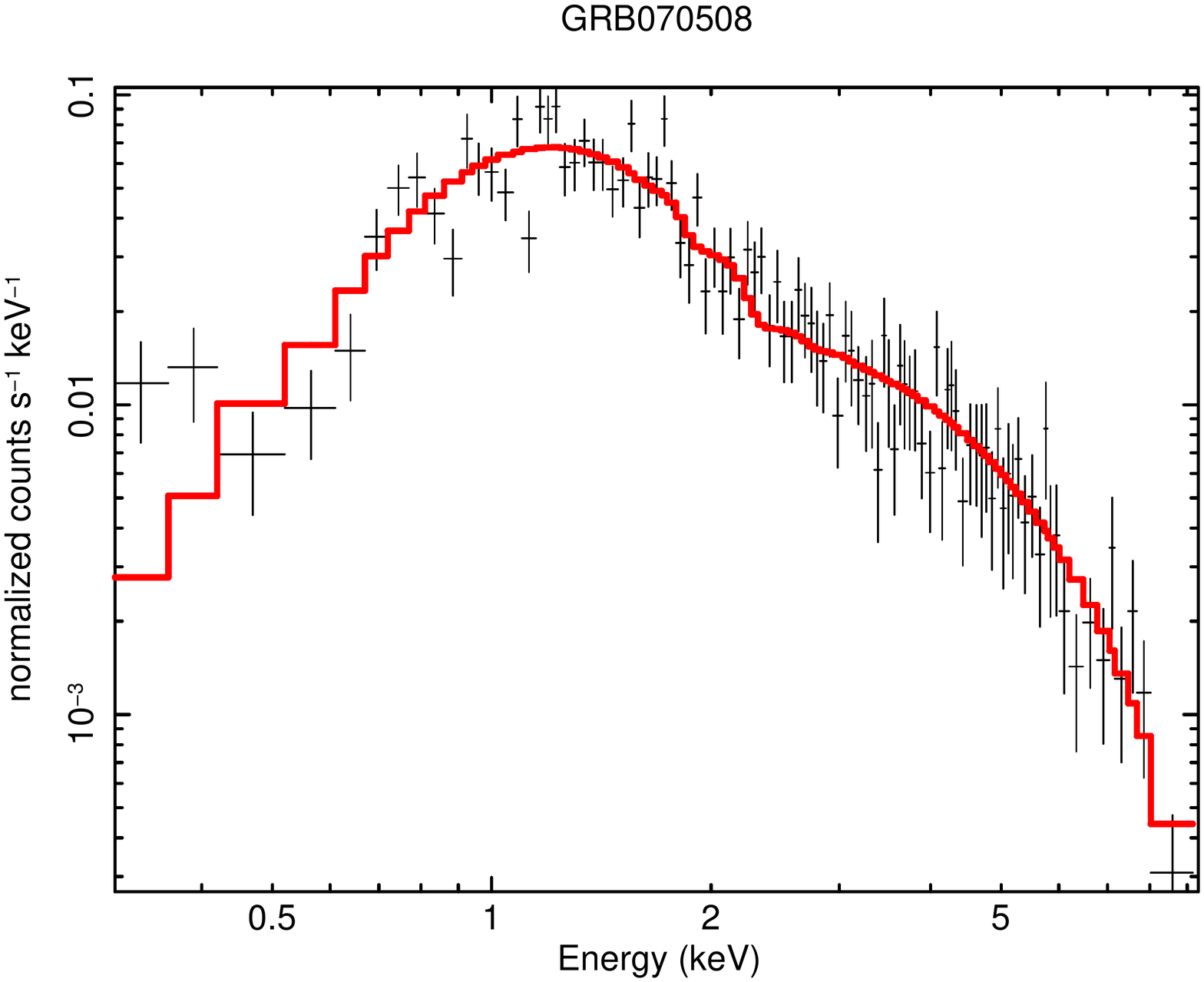}}
	\hfill
	\subfigure{\includegraphics[width=8.4cm, trim= 100 0 0 0]{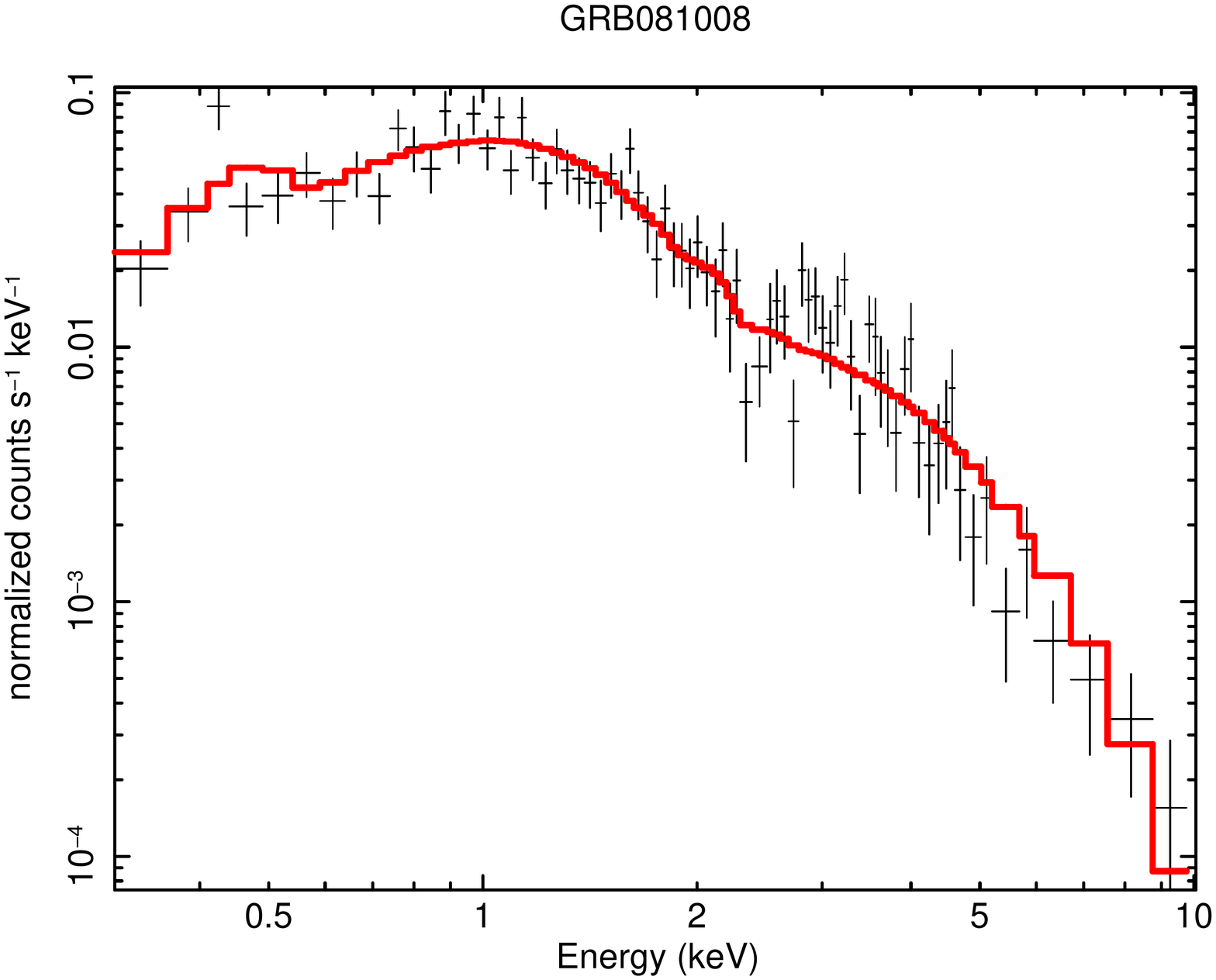}}
	\hfill
	\subfigure{\includegraphics[width=8.4cm, trim= 0 0 100 0]{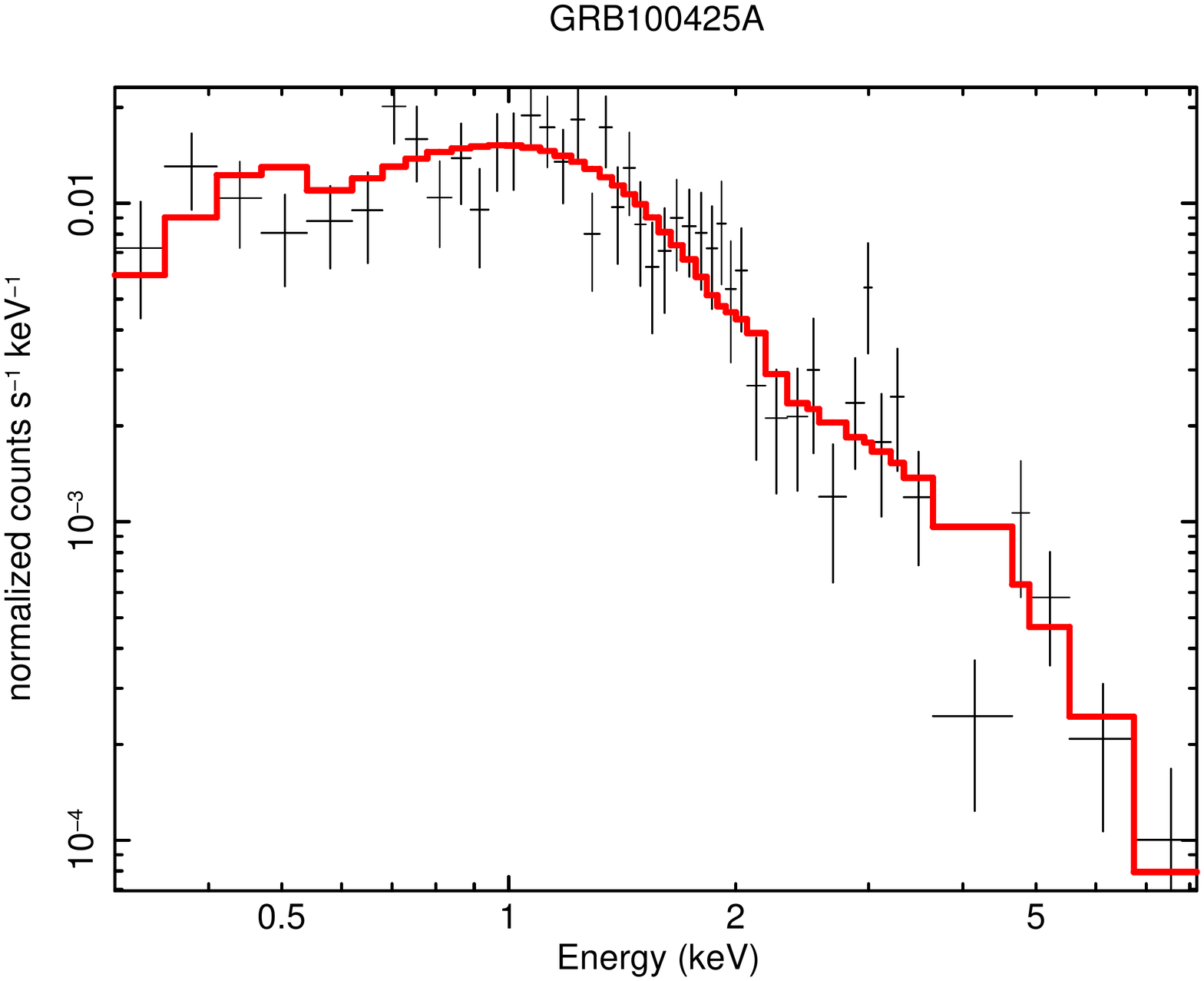}}
	\hfill
	\subfigure{\includegraphics[width=8.4cm, trim= 60 0 40 0]{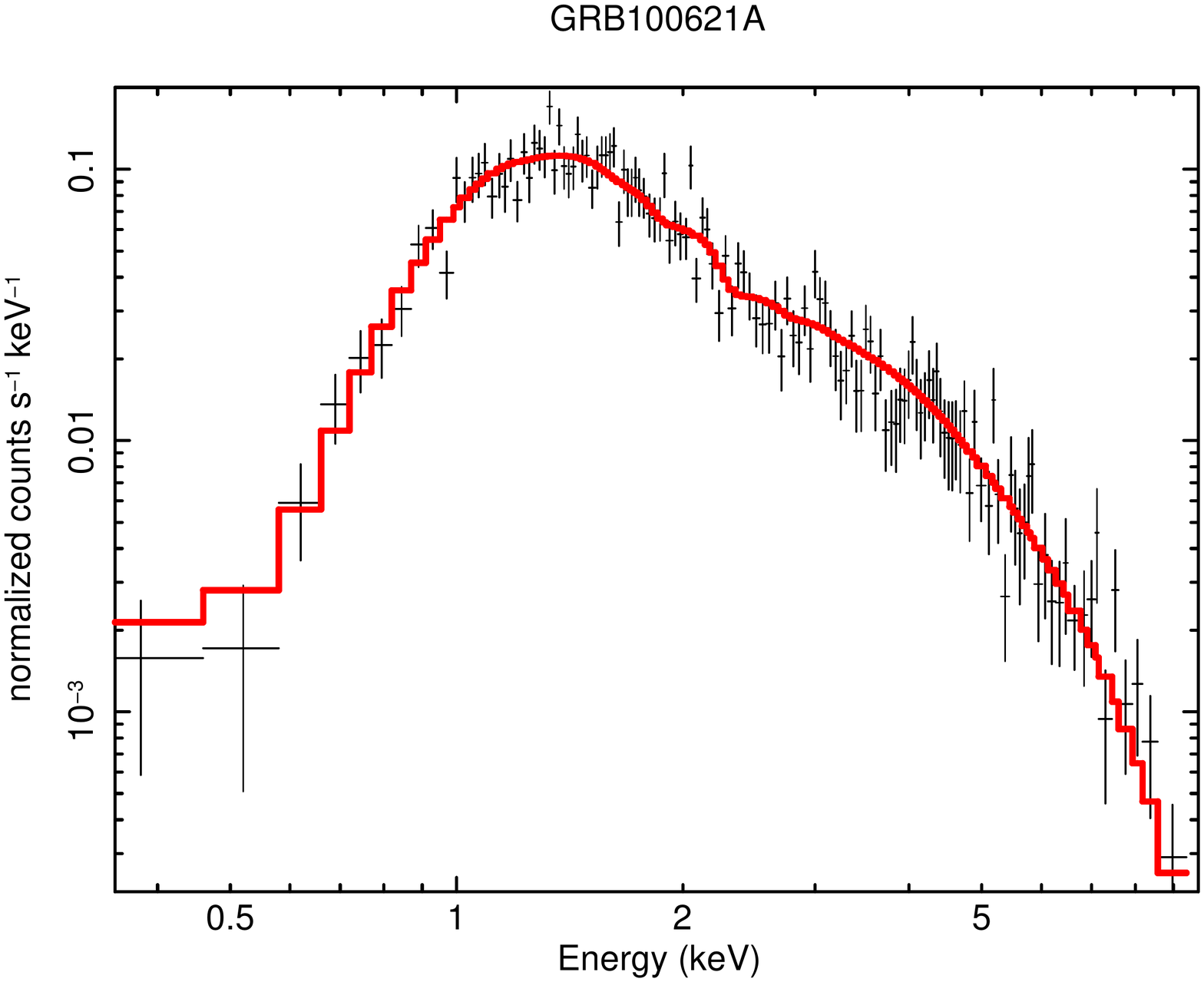}}
	\hfill
	\caption{Time averaged X-ray spectra for the four GRBs. The red curve shows the fit with XSpec using the new ATCA+GASS Galactic foreground data.}
	\label{fig:xspec_atca}
\end{figure*}

For our analysis we use the PC mode time averaged spectra and keep most of the initial parameters from the classical automated method. We use the same cosmology settings and the same ``cflux $\times$ TBabs $\times$ zTBabs $\times$ powerlaw'' theoretical model, where TBabs refers to the Galactic absorption and zTBabs refers to the intrinsic absorption from the host. The differences are the used Galactic absorption and longer iteration to achieve a more accurate fit \citep{racz2017,racz2018}. We compare the fitted intrinsic column densities for different Galactic foreground $N{\rm (HI)}$ values from the LAB survey, HI4PI, ATCA+GASS and three different calculations based on Planck data. Our results are in Tab.~\ref{tab:xspec}, where the intrinsic values are the fitted extragalactic hydrogen absorption column densities. The results show that the intrinsic density depends heavily on the correct estimation of the Galactic interstellar medium. We find that the Planck data, in particular the $N{\rm (H)}_{\rm DL}$, tends to give higher Galactic column densities which result in very low intrinsic column densities. The ATCA+GASS results for GRB081008 and GRB100425A are lower compared to the LAB results. However, considering the 90\% confidence of the fits all results agree with each other within the errors. Fig.~\ref{fig:xspec_atca} shows the fitted X-ray spectra using the Galactic hydrogen-absorption from our ATCA+GASS data. 

We note, that even though it is possible for the intrinsic column density to change over time due to ionisation from the GRB, \cite{racz2018} did not find any significant evidence for this in the XRT data considering the uncertainties of the measurement and the modelling. 	
	
\begin{table*}
	\centering
	\caption{
	Intrinsic hydrogen column density from Swift X-ray spectra with {\sc Xspec}. The upper and lower error limits show the 90\% confidence region. The columns show the intrinsic $N{\rm (H)}$ of the 4 GRBs assuming different foregrounds.}
	\label{tab:xspec}
	\begin{tabular}{l c c c c c c}
	\hline
	Name & $N(\mathrm{H})_\mathrm{LAB}$ & $N(\mathrm{H})_\mathrm{HI4PI}$ & $N(\mathrm{H})_\mathrm{ATCA}$ & $N(\mathrm{H})_\mathrm{PR1}$ & $N(\mathrm{H})_\mathrm{\rm RQ}$ & $N(\mathrm{H})_\mathrm{DL}$ \\
	& [$10^{22}$ cm$^{-2}$] &[$10^{22}$ cm$^{-2}$]  & [$10^{22}$ cm$^{-2}$]  &[$10^{22}$ cm$^{-2}$]   &[$10^{22}$ cm$^{-2}$] &[$10^{22}$ cm$^{-2}$]\\
	\hline
	GRB070508  & $0.94_{0.72}^{1.21}$ & $1.00^{1.26}_{0.76}$ & $0.93^{1.19}_{0.70}$ & $0.95^{1.22}_{0.73}$ & $0.88^{1.15}_{0.66}$ & $0.48^{0.72}_{0.26}$ \\
	GRB081008  & $0.32_{0.01}^{0.65}$ & $0.34^{0.68}_{0.03}$ & $0.13^{0.47}_{0.00}$ & $0.23^{0.56}_{0.00}$ & $0.30^{0.64}_{0.00}$ & $\approx0^{0.18}_{0.00}$ \\
	GRB100425A & $0.18_{0.00}^{0.64}$ & $0.19^{0.65}_{0.00}$ & $0.02^{0.49}_{0.00}$ & $0.02^{0.49}_{0.00}$ & $\approx0^{0.41}_{0.00}$ & $\approx0^{0.14}_{0.00}$ \\
	GRB100621A & $2.78_{2.51}^{3.06}$ & $2.79^{3.08}_{2.52}$ & $2.78^{3.06}_{2.51}$ & $2.79^{3.08}_{2.52}$ & $2.80^{3.09}_{2.53}$ & $2.74^{3.02}_{2.47}$ \\  
	\hline
	\end{tabular}
\end{table*}
	
\subsection{Intrinsic column density estimates from the literature}

Two of the GRBs from our sample have previous intrinsic column density estimates in the literature.
	
\subsubsection{GRB081008}
Beside the X-ray spectra fitting, there is a column density estimate in the literature for the host galaxy of GRB081008, which is in agreement with our estimates. An associated host galaxy to this GRB was detected with the Gemini-South GMOS-South spectrograph at a redshift of $z = 1.967$ \citep{cucchiara2008a, Cucchiara2015}. \cite{delia2011} derived an intrinsic hydrogen column density of $0.13 \pm 0.03 \times 10^{22}$ cm$^{-2}$ from the UVES/VLT optical spectra of the afterglow. However, \cite{lyman2017} could not confirm the detection of the host galaxy with HST F160W imaging. Considering that this galaxy is at a relatively high redshift, some of the X-ray absorption could be due to intervening IGM clouds (e.g. \citealt{Campana2010, Campana2012, Starling2013}).   
	
	
\subsubsection{GRB100621A}
	
GRB100621A, at a redshift of $z=0.542$, is embedded in a host galaxy with a total stellar mass of $\log M_{*}=9.61$ as derived from the Spitzer-measured luminosity \citep{Perley2016}. Based on modelling the  X-ray afterglow light curve, \cite{dado2018} concluded that supernova-less GRBs, such as GRB100621A, are associated with the birth of a millisecond pulsar and should be surrounded by low density ISM, similar to the environment in globular clusters. \HI\ observations of Galactic globular clusters, have measured column densities on the order of $10^{18}$ cm$^{-2}$ (e.g. \citealt{vanLoon2006}). 

In contrast to this, we find a value of $\sim 3 \times 10^{22}$cm$^{-2}$ for the intrinsic column density, which is the highest in our sample, and in the average column density range for GRBs. Intrinsic column densities from X-ray absorption typically range between $10^{20}$ and $10^{23}$cm$^{-2}$ (e.g. \citealt{Campana2012, Starling2013}). GRB100621A is at the furthest away from the Galactic plane ($b=-40.8$) and has the smallest Galactic foreground emission in our sample. Accordingly, we do not find any significant differences between our various foreground column density estimates. 
	
\section{Statistical investigation of the pattern of foreground ISM}
\label{sec:Fluctuations}
	
The Galactic ISM has fluctuations on a range of scales, in a fractal pattern, with larger amplitude fluctuations on the larger scales. This typically leads to a power-law in the Fourier Transform domain of spatial power as a function of spatial frequency. For \HI\ this is shown in the Southern Galactic Plane Survey (SGPS), a similar combination of ATCA and Parkes data to $\sim$~100~arcsec resolution, in the Galactic Plane by \citet{Dickey2001}. We tried to also test this with our ATCA+GASS data for these high latitude regions with lower \HI\ column densities.
	
We have done this with a simple Fourier Transform (FT) analysis, but have concluded that the results are not reliable given our low \HI\ signal-to-noise data. In particular, with the relatively small number of ATCA interferometer baselines, this FT domain or u,v-plane has areas without measured data that must be filled in the deconvolution process. This deconvolution is incomplete, so the areas without measured data are only partly corrected, and so are likely to be biased. We have also tried an alternative method of averaging u,v-data in annuli ({\sc miriad} task {\sc uvamp}), but conclude that this too is unreliable due to low signal-to-noise. Due to the vector nature of the data, a vector average would be decorrelated but the scalar average is biased by an uncertain noise correction.
	
With both methods, we confirm that there is less power on smaller scales, corresponding to longer interferometer spacings, in a rough power law, but we do not trust the derived value of the power law slope, for the reasons above. We suggest that better statistical results for the high latitude \HI\ power spectrum may be obtained with the GASKAP project \citep{GASKAP} which will have a wider field of view and more antennas compared to the ATCA.

\subsection{Fluctuations in the Planck data}

In addition to the \HI\ data, we also analysed the FIR data for fluctuations. We Fourier transformed the Planck $N{\rm (H)}$ maps with routines written in IDL based on the standard IDL FFT routine with periodogram normalization \citep{press1992}. For all maps the highest spatial sampling frequency corresponds to the pixel size of the used Planck images ($225\times225$). The Nyquist limit that allows correct sampling (without aliasing) is half of that frequency (\citealt{press1992}, Chap. 12.1), i.e. the double pixel size of the array. Therefore, the upper limit of the spatial frequency considered for the Fourier power spectrum was set to the Nyquist-limit. Instead of averaging $P(f)$ fluctuation power values in annuli for each $f=|f|$ spatial frequency, we used all individual data points $(f,P(f)$ pairs) to derive the final power spectrum. The spectral slope $\alpha$ is derived by robust line fitting to all data points in the $\log(f) - \log(P)$ space. 
	
\begin{figure*}
		\centering
		\includegraphics[width=2.0\columnwidth]{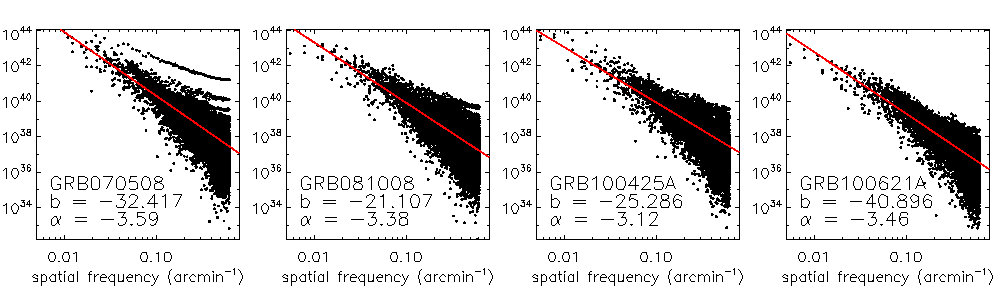}
		\caption{Power spectra of the target GRBs. Each dot represents a $(f,P(f))$ pair, the red line is the fitted linear. The high frequency end is determined by the Nyquist-limit, the low frequency end by the size of the map.}
		\label{fig:fir_stamp}
	\end{figure*}

Fig.~\ref{fig:fir_stamp} shows the power spectra of the Planck based hydrogen column densities of the four target regions with galactic latitudes and the spectral slopes. There is a moderate variation of the mean $N{\rm (H)}$ values of the fields ($1.276 \times 10^{21}$cm$^{-2}$, $7.395 \times 10^{20}$cm$^{-2}$, $8.849 \times 10^{20}$cm$^{-2}$, $2.435 \times 10^{20}$cm$^{-2}$), and we do not see a major difference in the spectral slopes either, although the values are typical for Galactic cirrus. We note here, that fields with higher hydrogen column densities usually have a steeper power spectrum as shown by earlier studies (e.g. \citealt{kiss2003}).
	
\section{Summary and Conclusions}
\label{sec:Summary}
	
Studying the afterglows of GRBs can give us valuable insights in the properties of the GRBs' host galaxy. For example, modelling the X-ray absorption of the afterglow can revile the intrinsic \HI\ content of the host galaxy. However, to get an accurate estimate of the intrinsic \HI\ content, we need to have precise measurements of the foreground Galactic \HI. Generally, LAB data is used for the Galactic foreground, nonetheless the spatial resolution of the LAB data is 0.6$^{\circ}$, which is significantly worse compared to the 18" resolution of the X-ray data. In this paper we investigate the effect of using single dish versus interferometric data to determine the foreground hydrogen column density. To this end, we present new high-resolution \HI\ observations with the ATCA towards 4 GRBs. We combine the interferometric ATCA data with single dish data from the Parkes telescope and derive new Galactic \HI\ column density values towards the GRBs at a spatial resolution of a few arc minutes. 

In addition to the spatial resolution, another important factor to consider for the Galactic foreground calculations is the presence of optically thick, cold \HI. Usually it is assumed that all the \HI\ is optically thin, however, this is not necessarily the case for many LOS. To investigate the optical depth of the foreground, We construct 1.4 GHz continuum maps from the ATCA observations and search for \HI\ absorption against continuum sources near the position of the GRBs. We detect ten continuum sources and we detect \HI\ absorption against two of them. We derive the optical depth spectra for these two lines of sight and correct the foreground column densities for two of the sources. We find that the optical depth correction substantially increases $N{\rm (HI)}$ for GRB070508 (60\%) and marginally increases $N{\rm (HI)}$ for GRB081008 (5\%). 

Complementary to the \HI\ data, we also derive the Galactic foreground column density from Planck FIR data. Comparing the results from the \HI\ and the FIR data, we find that the column densities agree relatively well. The exception is the column densities derived from $A_{\rm V}({\rm DL})$, which are about a factor of two higher compared to all the other column densities. We also find that the uncertainties for $N{\rm (H)}$ are the smallest for the \HI\ data. 

Using the new foreground column densities, we derive more accurate intrinsic hydrogen column density values for the GRB host galaxies. The results show that the intrinsic density does depend on the correct estimation of the Galactic interstellar medium. We find that higher Galactic column densities calculated from the interferometric and the FIR data result in lower intrinsic column densities. In particular, the ATCA+GASS intrinsic column densities for GRB081008 and GRB100425A are lower compared to the LAB results. However, considering the uncertainty of the X-ray spectral fitting, all results are consistent with each other. 

High angular resolution Galactic \HI\ maps show many small scale features, such as filaments and clouds, throughout the Galaxy. These maps can improve on the Galactic foreground used in GRB spectral fitting, which then provides better estimates on the intrinsic properties of the GRB host galaxy. We find that the ATCA+Parks maps have clearly the best spatial resolution and that the interferometric data can also be used to estimate the optical depth of the foreground. Upcoming surveys with new radio telescopes, such as the Galactic Australian SKA Pathfinder Survey (GASKAP; \citealt{GASKAP}) with the Australian Square Kilometre Array Pathfinder (ASKAP) and possible surveys with the Square Kilometer Array (SKA) will provide Galacitic \HI\ data with arcsecond scale resolution which will largely improve on the foreground estimation of GRBs. Until this data is available, we recommend the use of the better resolution and sensitivity HI4PI data over the traditionally used LAB data.
	
\section*{Acknowledgements}
	
The Australia Telescope Compact Array and the Parkes telescope are part of the Australia Telescope National Facility which is funded by the Commonwealth of Australia for operation as a National Facility managed by CSIRO. Based on observations obtained with Planck (http://www.esa.int/Planck), an ESA science mission with instruments and contributions directly funded by ESA Member States, NASA, and Canada. Useful discussions with Csaba Kiss are highly appreciated. We would like to thank the anonymous referee for the useful comments and suggestions, which helped us to improve the paper. This research was partly supported by the OTKA grants K101393 and NN-111016, the Hungarian NKFI TKP grant, the NAOJ ALMA Scientific Research Grant Number 2016-03B, and the Science Research Grants of the Japan Society for the Promotion of Science (25247016, 18H01250). Supported through the \'UNKP-17-3 New National Excellence Program of the Hungarian Ministry of Human Capacities. This research has made use of data supplied by the UK Swift Science Data Centre at the University of Leicester, NASA's Astrophysics Data System, {\sc matplotlib} \citep{Hunter2007}, {\sc NumPy} \citep{vanderWalt2011}, and {\sc Astropy}, a community-developed core {\sc Python} package for Astronomy (\citealt{Astropy}; http://www.astropy.org).
	
	
	
\bibliographystyle{mnras}
\bibliography{GRB_paper} 
	
	
\FloatBarrier
\appendix
\section{Large scale environment of the targeted regions}
\label{appendix:large_scale}
	
\begin{figure*}
	\includegraphics[width=16cm]{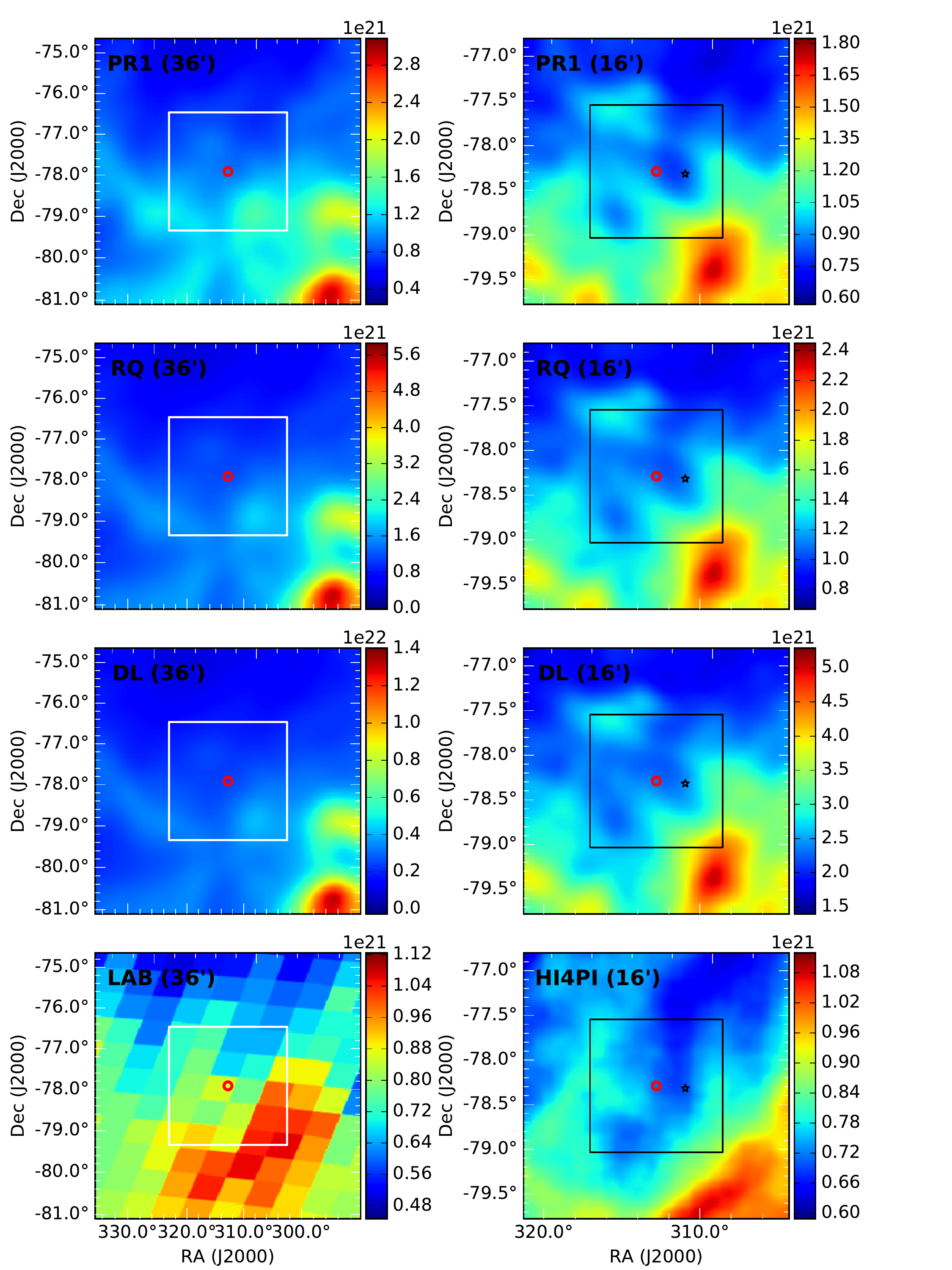}
	\caption{Planck, LAB and HI4PI $N($H$)$ maps toward GRB070508. Left column: The first three images from top are Planck PR1, Planck PR2 RQ, and Planck PR2 DL based $N($H$)_\mathrm{MW}$ distributions smoothed to 36' resolution; the bottom one is the $N($H$)_\mathrm{MW}$ from the LAB survey. Right column: All images show the central area of the left column (marked with white box), but with a resolution of 16'. The top three rows show the same data on the left and the right hand side, while the bottom row shows HI4PI data on the right side. The black box indicates the area where the ATCA measurements were taken, the red circle indicates the position of the GRB and the black asterisk denotes the position of the continuum source.}
	\label{fig:GRB070508_maps}
\end{figure*}

\begin{figure*}
	\includegraphics[width=16cm]{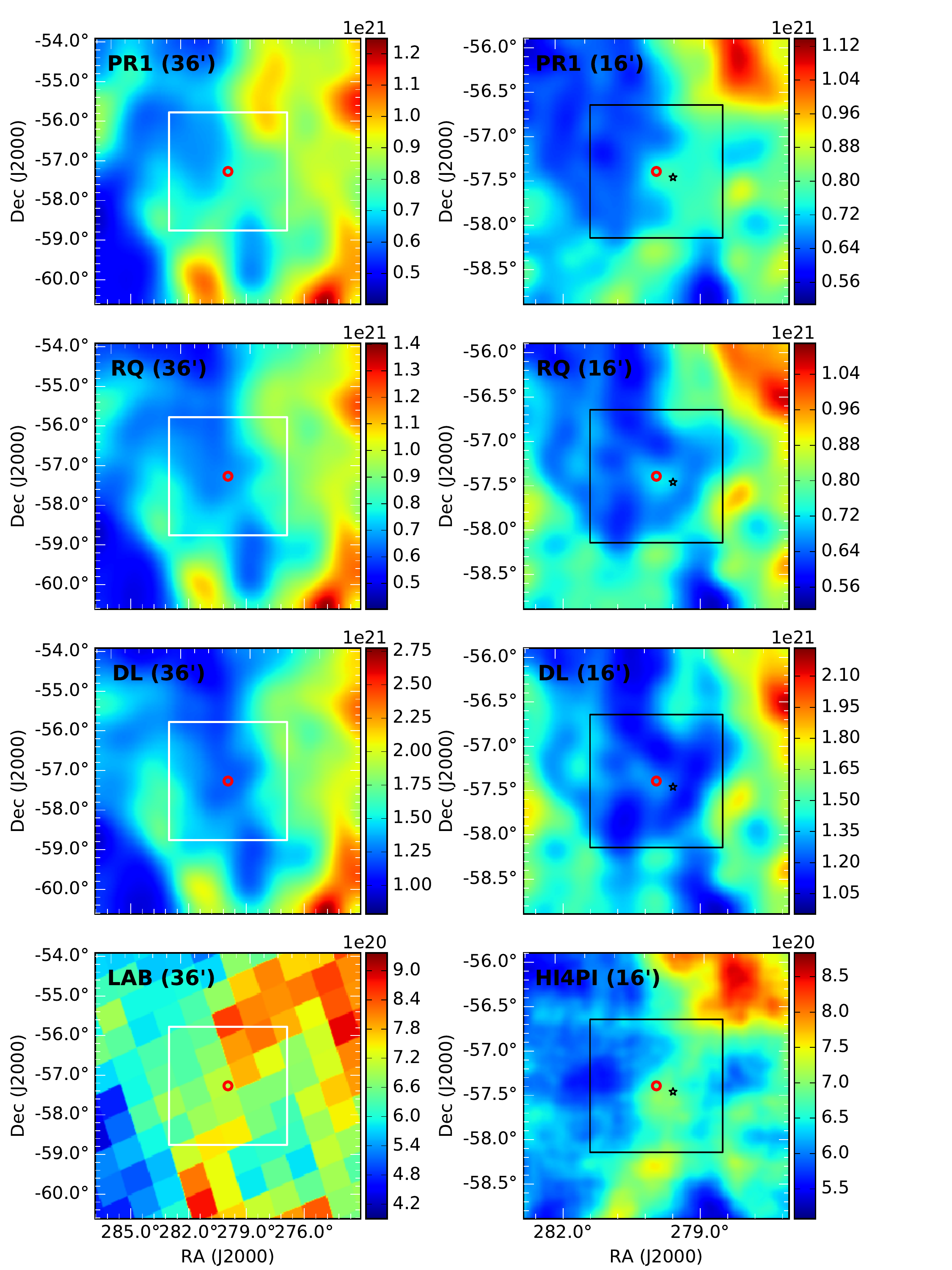}
	\caption{Same as Fig.\ref{fig:GRB070508_maps} in the direction of GRB081008.}
	\label{fig:GRB081008_maps}
\end{figure*}

\begin{figure*}
	\includegraphics[width=16cm]{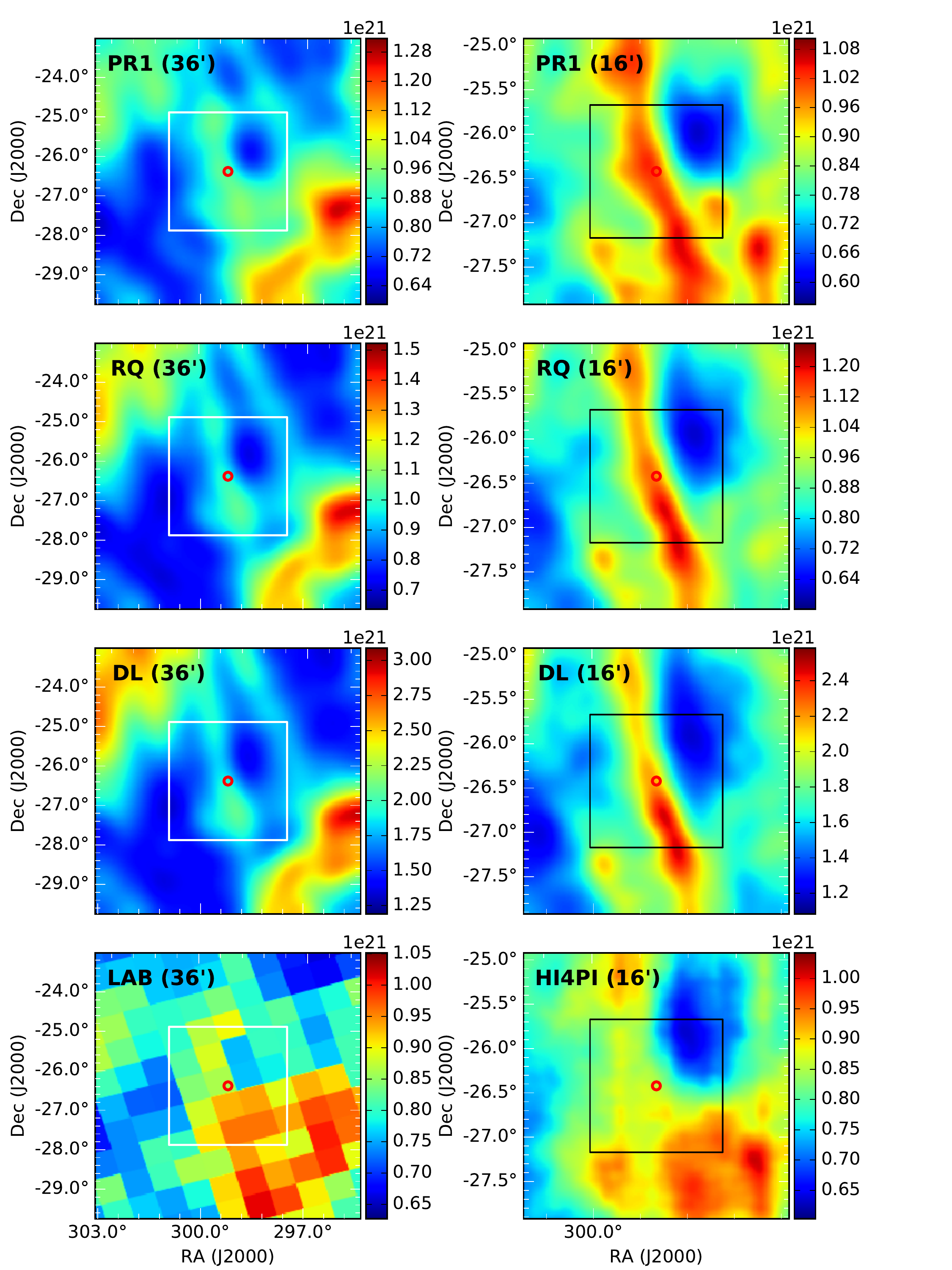}
	\caption{Same as Fig.\ref{fig:GRB070508_maps} in the direction of GRB100425.}
	\label{fig:GRB100425_maps}
\end{figure*}

\begin{figure*}
	\includegraphics[width=16cm]{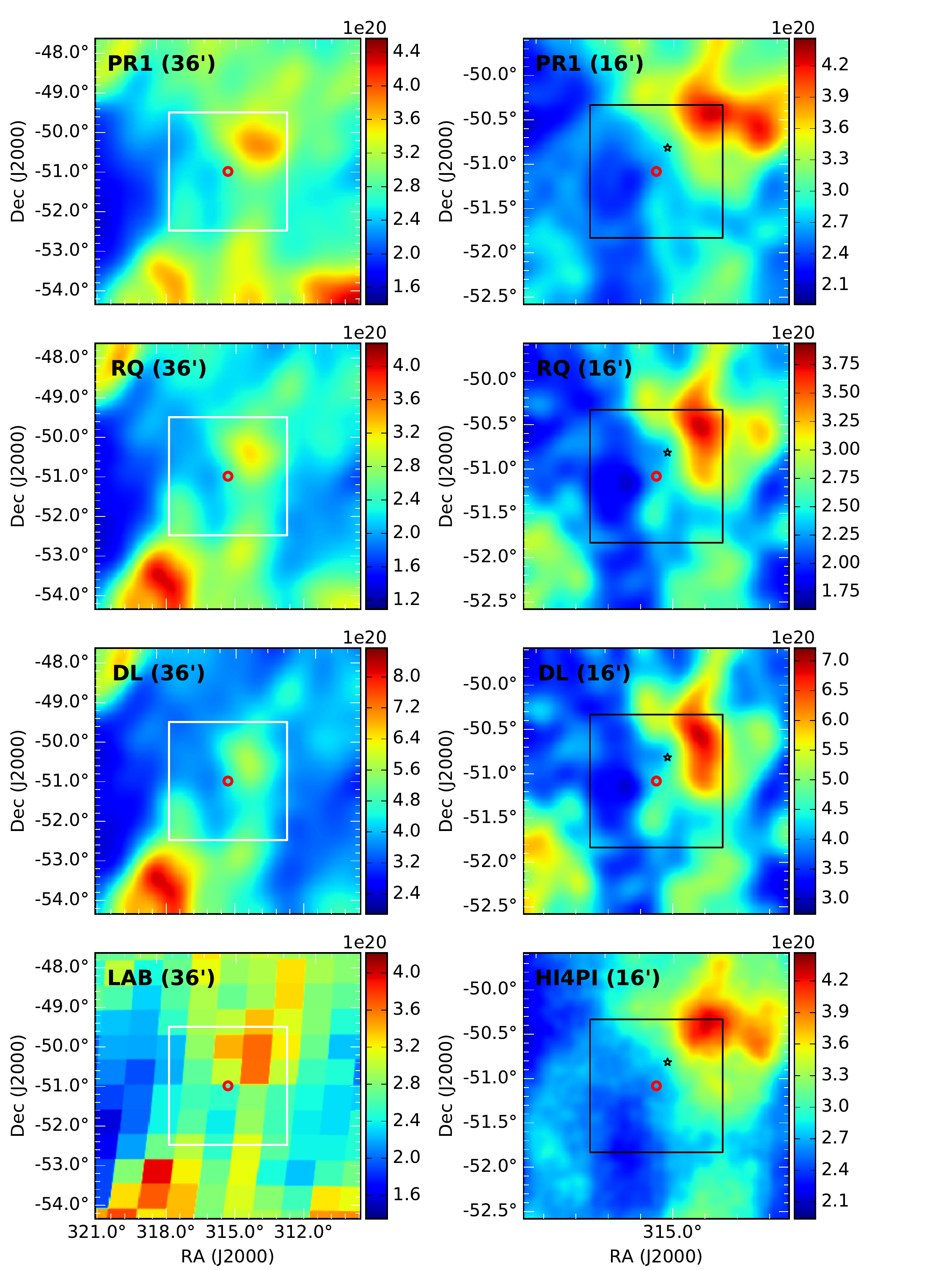}
	\caption{Same as Fig.\ref{fig:GRB070508_maps} in the direction of GRB100621.}
	\label{fig:GRB100621_maps}
\end{figure*}

\begin{figure*}
	\includegraphics[width = 0.8\textwidth]{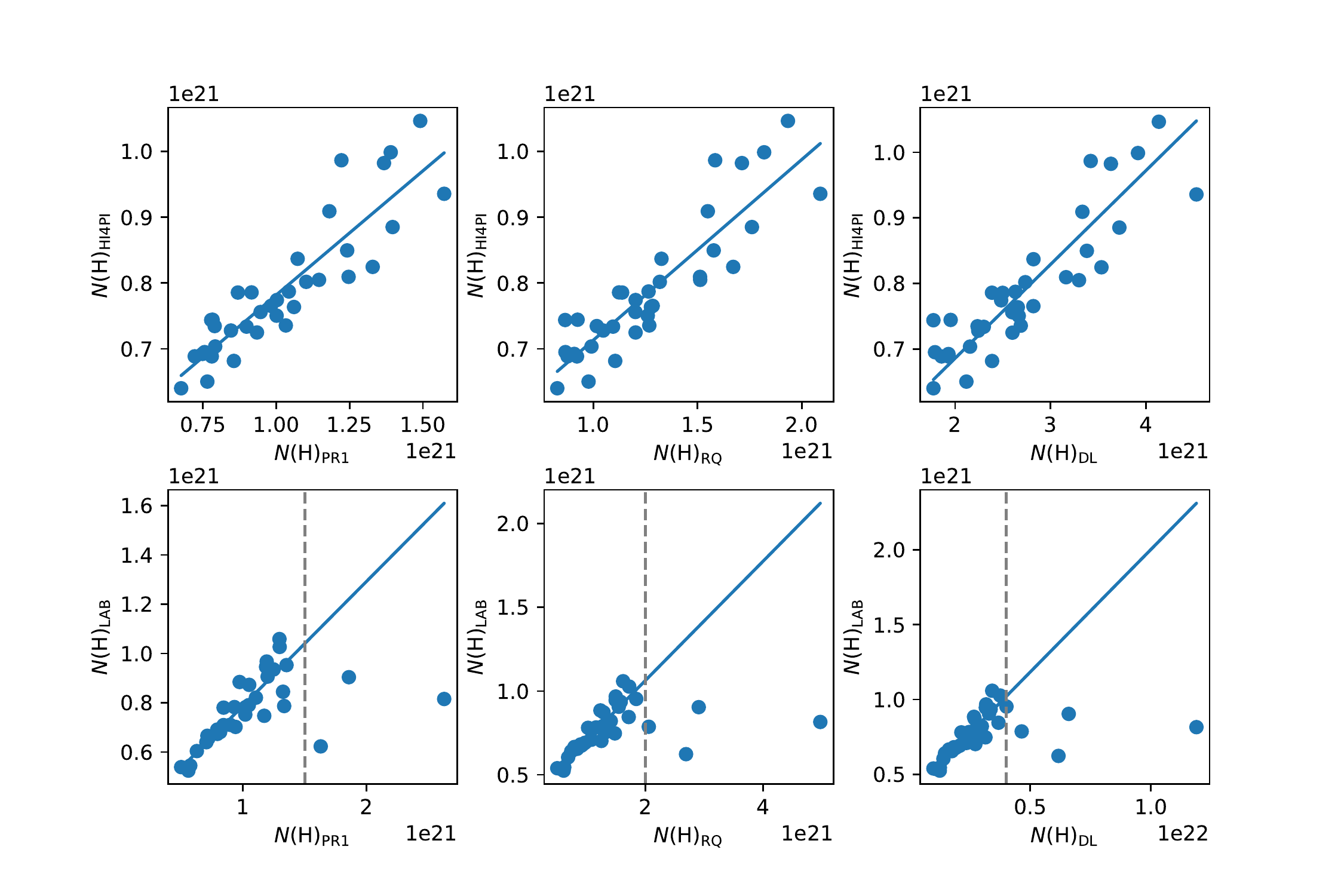}
	\caption{Correlation plots of the estimated Galactic foreground hydrogen column densities in the direction of {GRB070508}, based on various all sky surveys. Top row from left to right: HI4PI vs. Planck PR1; HI4PI vs. Planck PR2 RQ; HI4PI vs. Planck PR2 DL. The bottom row: LAB survey vs. Planck PR1; LAB survey vs. Planck PR2 RQ; LAB survey vs. Planck PR2 DL. Ordinary least square fitted lines are drawn. The Pearson's correlation coefficients and the least square fits were calculated only at the linear parts of the correlation plots with HI data, the limits are indicated with dashed vertical lines.}
	\label{fig:GRB070508_correlation}
\end{figure*}

\begin{figure*}
	\includegraphics[width = 0.9\textwidth]{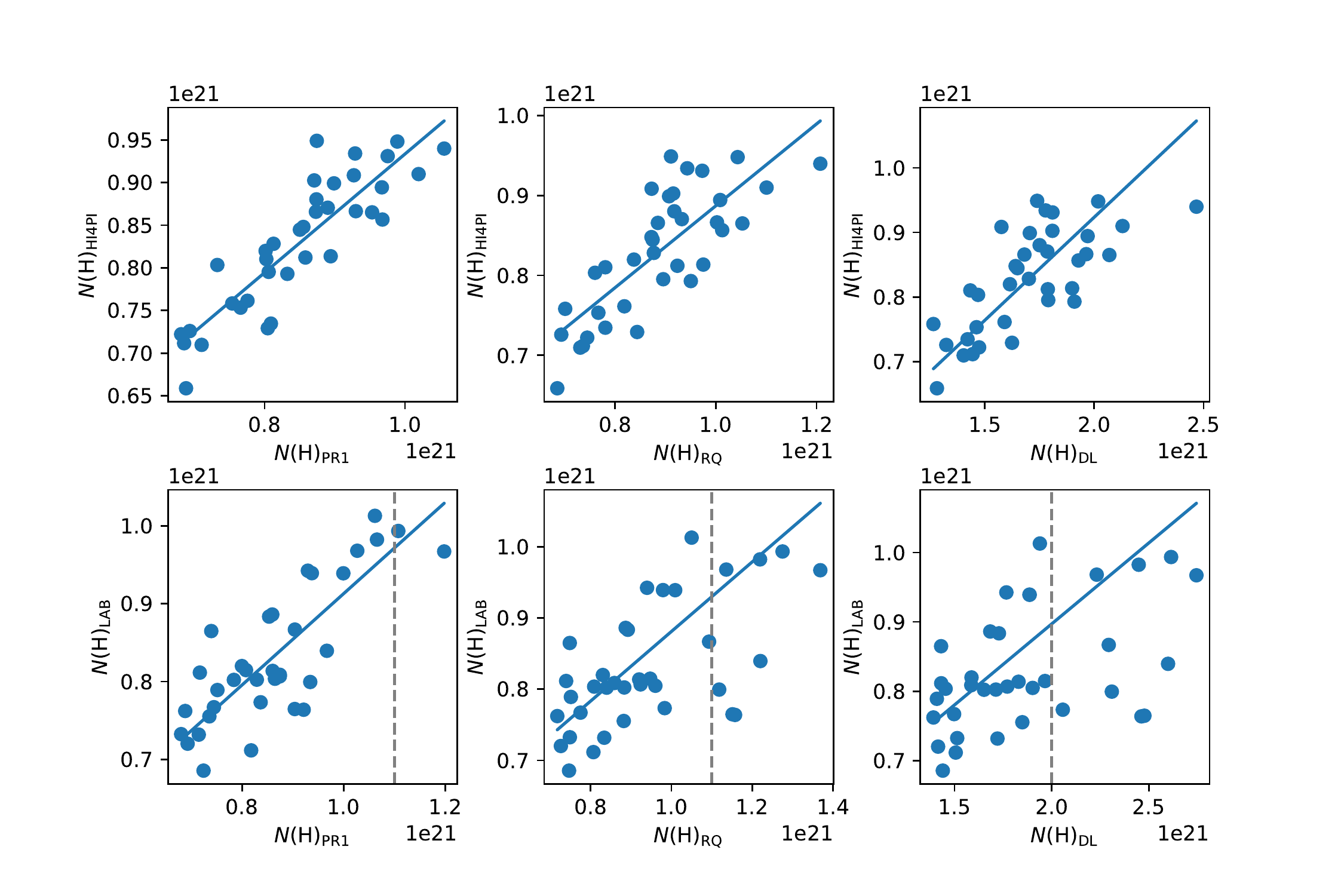}
	\caption{Same as Figure \ref{fig:GRB070508_correlation} for GRB100425.}
	\label{fig:GRB100425_correlation}
\end{figure*}

\begin{figure*}
	\includegraphics[width = 0.9\textwidth]{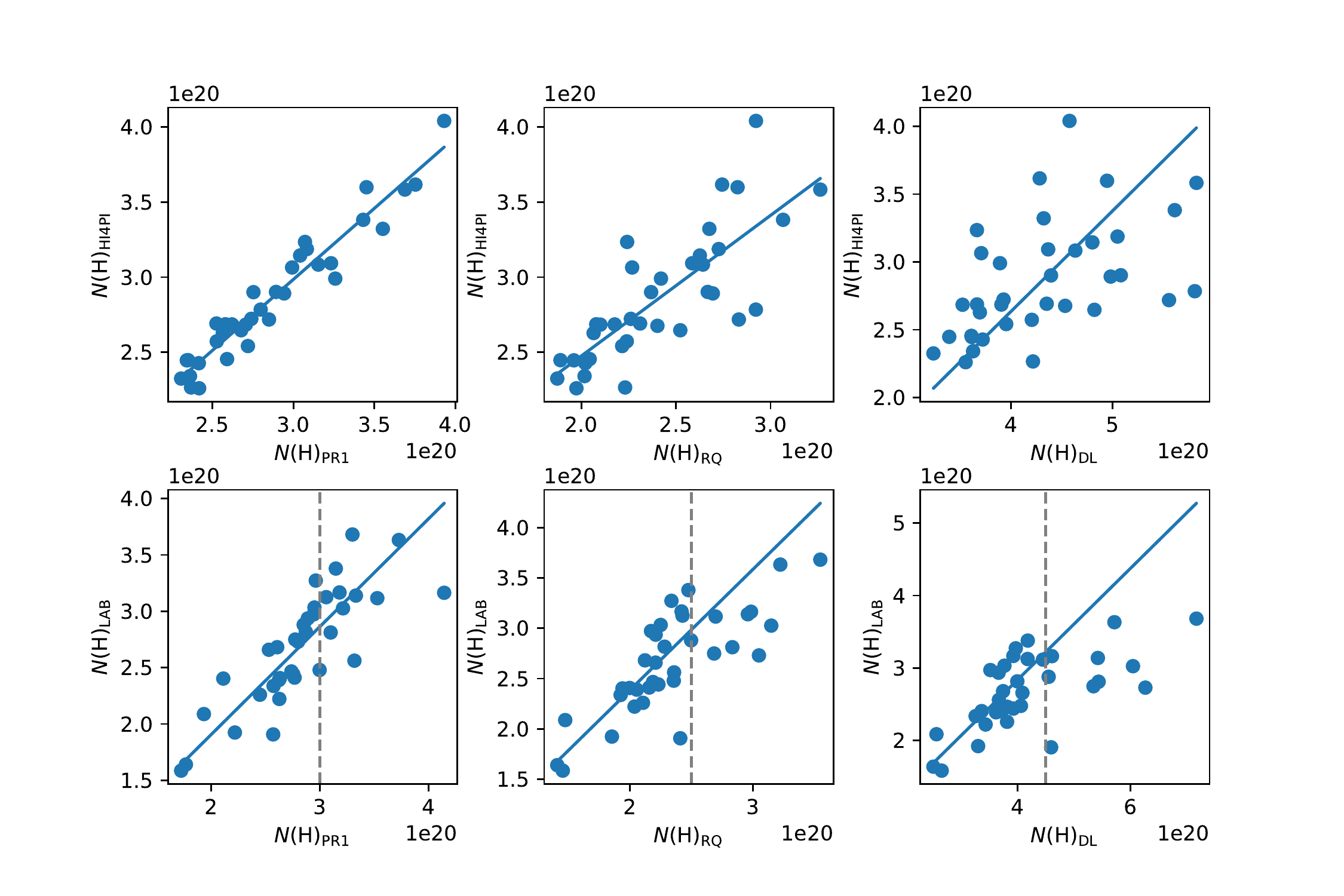}
	\caption{Same as Figure \ref{fig:GRB070508_correlation} for GRB100621.}
	\label{fig:GRB100621_correlation}
\end{figure*}

We investigate the large scale environment of the studied locations using various all sky surveys. In Fig.~\ref{fig:GRB070508_maps},~\ref{fig:GRB081008_maps},~\ref{fig:GRB100425_maps} and \ref{fig:GRB100621_maps} we compare hydrogen column density maps using Planck results by \cite{planck2014} and by \cite{planck_dust}, and column density maps from the LAB Survey \citep{kalberla2005}, and the HI4PI survey \citep{hi4pi2016}. The left column of the figures shows a six square degree region around the GRBs and the right hand side a 3 square degree region. Fig.~\ref{fig:GRB070508_correlation},~\ref{fig:GRB081008_maps},~\ref{fig:GRB100425_correlation} and \ref{fig:GRB100621_correlation} show the correlation between the different maps. The blue solid line is the ordinary least square fit to the data and Tab.~\ref{tab:correlation_coefficients} summarises the correlation coefficients. 

Throughout the 4 regions we see an order of magnitude variation of $N{\rm (H)}$ and quite complex patterns. We find that the infrared data shows similar structures to the \HI\ data, but the shape and location of the highest intensity clumps tend to differ. Overall the PR1 Planck maps show the most similar features to the \HI\ maps and these two datasets have the tightest correlation in $N{\rm (H)}$.  

For example in the case of GRB100425A (Fig.~\ref{fig:GRB100425_maps}) there is a filamentary structure going trough the middle of the Planck images. In the \HI\ images, the main feature is a cloud like structure in the bottom right corner of the image and there is only a faint filament like structure, which is spatially offset from the infrared filament. In the case of GRB070508 (Fig.~\ref{fig:GRB070508_maps}) the 16' images have the same features, but the highest intensity regions in the LAB map are offset to the highest intensity regions of the Planck 36' maps. This is also shown by the offset points in Fig~\ref{fig:GRB070508_correlation}. The offset feature in these images suggest that some of the brighter FIR regions are not overlapping with, but rather surrounded by or enveloped by \HI.     

\begin{table*}
	\centering
	\caption{Correlation coefficients between the \HI\ column densities measured by the different \HI\ and FIR surveys.}
	\label{tab:correlation_coefficients}
	\begin{tabular}{l c c c c c c}
	\hline
	Name & HI4PI-PR1 & HI4PI-RQ & HI4PI-DL & LAB-PR1 & LAB-RQ & LAB-DL\\
	\hline
	GRB081008 & 0.92 & 0.72 & 0.48 & 0.85 & 0.78 & 0.59\\  
	GRB100425 & 0.88 & 0.79 & 0.72 & 0.77 & 0.65 & 0.56\\
	GRB100621 & 0.96 & 0.79 & 0.51 & 0.83 & 0.76 & 0.79\\
	GRB070508 & 0.89 & 0.89 & 0.89 & 0.9 & 0.92 & 0.91\\
	\hline
	\end{tabular}
\end{table*}
	
\begin{figure*}
	\includegraphics[width = 0.9\textwidth]{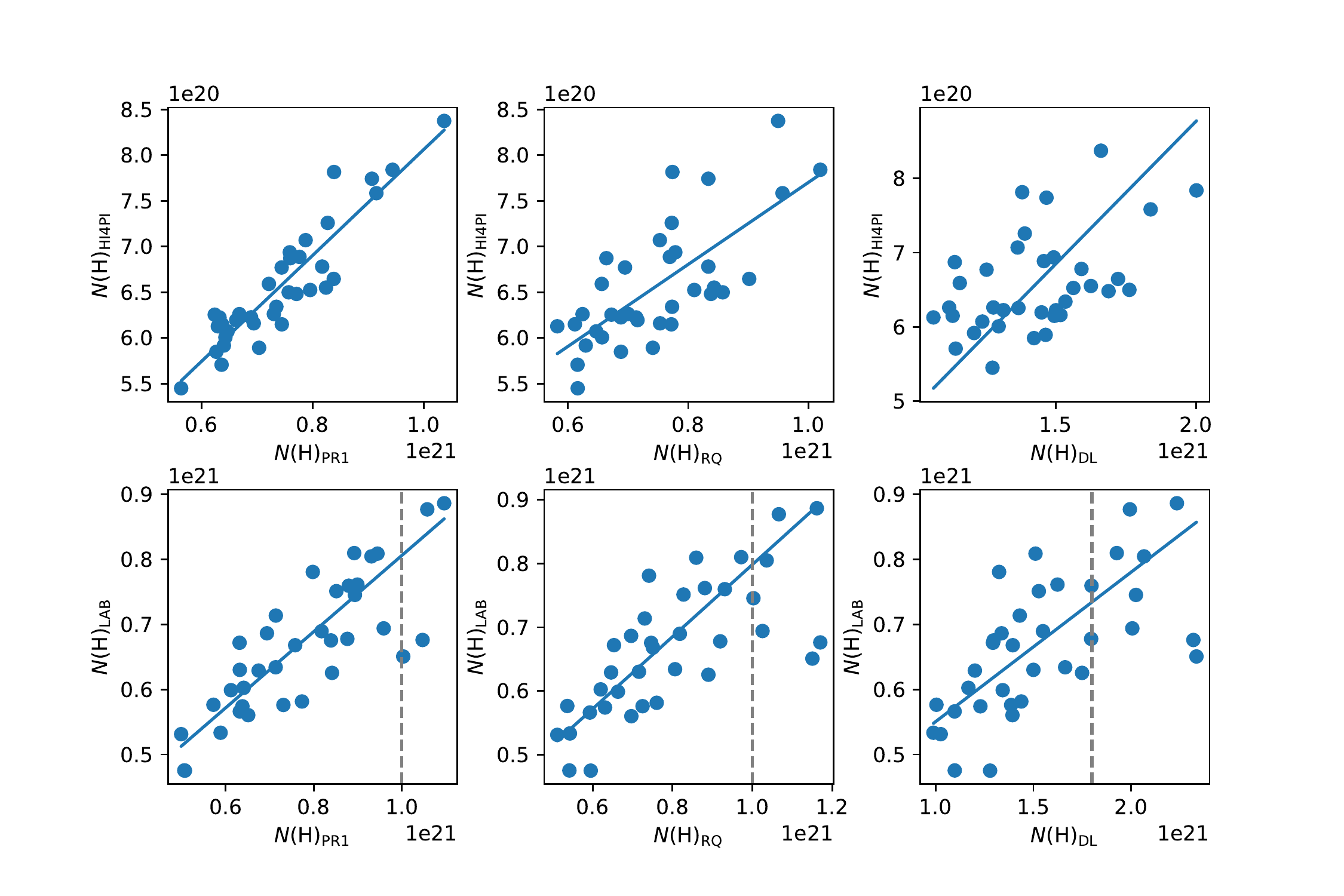}
	\caption{Same as Figure \ref{fig:GRB070508_correlation} for GRB081008.}
	\label{fig:GRB081008_correlation}
\end{figure*} 
	
\label{lastpage}
\end{document}